\documentclass[preprint,12pt]{elsarticle}
\usepackage[a4paper, left=1.1in,right=1.1in,top=1in,bottom=1.1in]{geometry}
\usepackage{lineno}
\usepackage{hyperref}
\usepackage{url}
\usepackage{epsfig}
\usepackage{xcolor}
\usepackage{textcomp}
\usepackage{mismath}
\modulolinenumbers[5]

\journal{Nuclear Instruments and Methods in Physics Research Section A}

\begin{document}

\begin{frontmatter}

\title{BINGO innovative assembly for background reduction in bolometric $0\nu\beta\beta$ experiments} 

\author[1]{A.~Armatol}
\author[2]{C.~Augier}
\author[b]{I.C.~Bandac}
\author[1]{D.~Baudin}
\author[3]{G.~Benato}
\author[1]{V.~Berest}
\author[5]{L.~Berg{\'e}}
\author[2]{J.~Billard}
\author[b,f,g]{J.M.~Calvo-Mozota}
\author[4]{P.~Carniti}
\author[5]{M.~Chapellier}
\author[7,8]{F.A.~Danevich}
\author[2]{M.~De Jesus}
\author[1,5]{T.~Dixon}
\author[5]{L.~Dumoulin}
\author[1]{F.~Ferri}
\author[2]{J.~Gascon}
\author[5]{A.~Giuliani}
\author[1]{H.~Gomez}
\author[4]{C.~Gotti}
\author[1]{Ph.~Gras}
\author[1]{M.~Gros}
\author[2]{A.~Juillard}
\author[1]{H.~Khalife\corref{correspondingauthor}}
\ead{hawraa.khalife@cea.fr}
\author[7]{V.V.~Kobychev}
\author[2]{H.~Lattaud}
\author[1]{M.~Lefevre}
\author[5]{P.~Loaiza}
\author[5]{P.~de~Marcillac}
\author[5]{S.~Marnieros}
\author[5]{C.A.~Marrache-Kikuchi}
\author[n,o]{M.~Martinez}
\author[1]{Ph.~Mas}
\author[1]{E.~Mazzucato}
\author[1]{J.F.~Millot}
\author[1]{C.~Nones}
\author[5]{E.~Olivieri}
\author[n]{A.~Ortiz de Sol\'orzano}
\author[4]{G.~Pessina}
\author[5]{D.V.~Poda\corref{correspondingauthor}}
\ead{denys.poda@ijclab.in2p3.fr}
\author[p]{A.~Rojas}
\author[5]{J.A.~Scarpaci}
\author[1]{B.~Schmidt}
\author[1]{O.~Tellier}
\author[3,7]{V.I.~Tretyak}
\author[p]{G.~Warot}
\author[p]{Th.~Zampieri} 
\author[7]{M.M.~Zarytskyy}
\author[1]{A.~Zolotarova}

\address[1]{IRFU, CEA, Université Paris-Saclay, F-91191 Gif-sur-Yvette, France}
\address[2]{Institut de Physique des 2 Infinis, Villeurbanne, France}
\address[b]{Laboratorio Subterr\'aneo de Canfranc, 22880 Canfranc-Estaci\'on, Spain}
\address[3]{INFN Laboratori Nazionali del Gran Sasso, I-67100 Assergi (AQ), Italy}
\address[5]{Universit\'e Paris-Saclay, CNRS/IN2P3, IJCLab, 91405 Orsay, France}
\address[f]{Escuela Superior de Ingenier\'ia, Ciencia y Tecnolog\'ia, Universidad Internacional de Valencia -- VIU, 46002 Valencia, Spain}
\address[g]{Escuela Superior de Ingenier\'ia y Tecnolog\'ia, Universidad Internacional de La Rioja, 26006 Logro\~no, Spain}
\address[4]{INFN Sezione di Milano-Bicocca, I-20126 Milano, Italy}
\address[7]{Institute for Nuclear Research of NASU, 03028 Kyiv, Ukraine}
\address[8]{INFN Sezione di Roma Tor Vergata, Rome, Italy}
\address[n]{Centro de Astropart\'iculas y F\'isica de Altas Energ\'ias, Universidad de Zaragoza, 50009 Zaragoza, Spain}
\address[o]{ARAID Fundaci\'on Agencia Aragonesa para la Investigaci\'on y el Desarrollo, 50018  Zaragoza, Spain}
\address[p]{Univ. Grenoble Alpes, CNRS, Grenoble INP, LPSC/LSM-IN2P3, 73500 Modane, France}

\cortext[correspondingauthor]{Corresponding authors}

\begin{abstract}
BINGO is a project aiming to set the grounds for large-scale bolometric neutrinoless double-beta-decay experiments capable of investigating the effective Majorana neutrino mass at a few meV level. It focuses on developing innovative technologies (a detector assembly, cryogenic photodetectors and active veto) to achieve a very low background index, of the order of $10^{-5}$~counts/(keV~kg~yr) in the region of interest. The BINGO demonstrator, called MINI-BINGO,  is designed to investigate the promising double-beta-decay isotopes $^{100}$Mo and $^{130}$Te and it will be composed of Li$_2$MoO$_4$ and TeO$_2$ crystals coupled to bolometric light detectors and surrounded by a Bi$_4$Ge$_3$O$_{12}$-based veto. This will allow us to reject a significant background in bolometers caused by surface contamination from $\alpha$-active radionuclides by means of light yield selection and to mitigate other sources of background, such as surface contamination from $\beta$-active radionuclides, external $\gamma$ radioactivity, and pile-up due to random coincidence of background events. This paper describes an R\&D program towards the BINGO goals, particularly focusing on the development of an innovative assembly designed to reduce the passive materials within the line of sight of the detectors, which is expected to be a dominant source of background in next-generation bolometric experiments. We present the performance of two prototype modules --- housing four cubic (4.5-cm side) Li$_2$MoO$_4$ crystals in total --- operated in the Canfranc underground laboratory in Spain within a facility developed for the CROSS double-beta-decay experiment. 
\end{abstract}


\end{frontmatter}


\section{Introduction}

Neutrinoless double-beta decay ($0\nu\beta\beta$) is a Standard-Model-forbidden rare nuclear transition that can occur only if neutrinos are Majorana particles ($\nu=\bar\nu$) \cite{Schechter:1982}. It has been proposed by W. Furry in 1939 \cite{Furry:1939}, four years after M. Goeppert-Mayer suggested the existence of ordinary two-neutrino double-beta decay ($2\nu\beta\beta$) \cite{GoeppertMayer:1935}. In $2\nu\beta\beta$, a nucleus ($A, Z$) spontaneously decays into an isobar ($A, Z+2$) with the emission of two electrons and two anti-neutrinos, whereas in $0\nu\beta\beta$ we expect no emission of neutrinos. Consequently, its signature is a peak in the summed energy spectrum of the final state electrons at the $Q$-value of the decay ($Q_{\beta\beta}$). The $0\nu\beta\beta$ process would violate lepton number and requires physics beyond the Standard Model \cite{GomezCadenas:2023,Agostini:2022,Adams:2022b}. Moreover, from the observation of the $0\nu\beta\beta$ it could be inferred the effective Majorana mass term for the neutrino, if the decay is mediated by light Majorana neutrinos, as foreseen by the simplest extension of the Standard Model \cite{GomezCadenas:2023,Agostini:2022,Adams:2022b}.

There are several tens of nuclei capable of undergoing $2\nu\beta\beta$ and only 11 have been observed experimentally \cite{Pritychenko:2023}. Potentially, all these nuclei can undergo $0\nu\beta\beta$ too. However, only a few are considered as good candidates to search for $0\nu\beta\beta$ experimentally. This is related to the availability of a suitable detector technology, a low background in the region of interest (ROI), i.e. around $Q_{\beta\beta}$, a high isotopic abundance of the isotope of interest and/or the possibility of isotopic enrichment at large scale \cite{GomezCadenas:2023,Agostini:2022}. 

The sensitivity of a $0\nu\beta\beta$ search can be maximised by working on four factors: an ultra low background, a high energy resolution, a large $\beta\beta$ isotopic mass, and a long live-time. In this context, BINGO (Bi-Isotope $0\nu\beta\beta$ Next Generation Observatory) \cite{BINGO:webpage, Khalife:2023} is an ERC-funded project \cite{BINGO:ERC} that focuses on developing techniques to achieve an ultra low background for $0\nu\beta\beta$ experiments, exploiting one of the most promising detectors for $0\nu\beta\beta$ searches, cryogenic bolometers\cite{Fiorini:1983, Pirro:2017, Poda:2017, Bellini:2018, Biassoni:2020, Poda:2021, Zolotarova:2021a}. Its purpose is to set the grounds for future large-scale bolometric searches \cite{cupid1t:2022}.

In this article, we will briefly overview the BINGO project and will present in detail one of the three main lines of its R\&D program, the most crucial one in view of reducing backgrounds induced by detector structure components. Particularly, we will focus on the design of the innovative detector assembly and on the results from the first cryogenic test of two BINGO modules with four large-volume ($\sim$90~cm$^3$) crystal scintillators, each coupled to a large-area ($\sim$20~cm$^2$) bolometric semiconductor-based photodetector.

\section{BINGO concept for high-sensitivity $0\nu\beta\beta$ decay searches}

The BINGO project is dedicated to explore new methods for the background reduction in experiments searching for $0\nu\beta\beta$ decay with thermal detectors. This can be achieved by the following actions:
\begin{itemize}
    \item Selection of two isotopes of interest embedded in detector materials suitable for the well-established technology of hybrid heat-light cryogenic bolometers, promising for multi-isotope $0\nu\beta\beta$ decay searches \cite{Giuliani:2018} thanks to detector material flexibility, high energy resolution and particle identification capability \cite{Poda:2017};
    
    \item Development of a novel detector assembly with a reduction of the passive materials facing the detector, leading to a substantial suppression of surface-induced background;
    
    \item Development of high-performance bolometric light detectors with thermal signal amplification, utilized for particle identification allowing for an active reduction of the background induced by surface $\alpha$ radioactivity;
    
    \item Development of an active shield, based on efficient scintillators with bolometric light detector readout to surround the experimental volume, enabling an active reduction of background induced by radioactive sources outside the detector;
    
    \item Demonstration of the innovative technologies in a small-scale experiment, necessary for the implementation of detector technologies at a large scale.
\end{itemize}
All these key ingredients of the BINGO R\&D program are briefly presented in this section.

\subsection{BINGO: towards bi-isotope bolometric $0\nu\beta\beta$ experiment with particle identification}

BINGO will study the two promising $0\nu\beta\beta$ isotopes $^{100}$Mo ($Q_{\beta\beta}=3034$~keV \cite{Wang:2017a}) and $^{130}$Te ($Q_{\beta\beta}=2527$~keV \cite{Wang:2017a}) embedded in the well known bolometric compounds Li$_2$MoO$_4$ (lithium molybdate) and TeO$_2$ (tellurium dioxide\footnote{There are two forms of TeO$_2$: synthetic ($\alpha$-TeO$_2$, paratellurite) and natural ($\beta$-TeO$_2$, tellurite). The synthetic form is typically used as a material for TeO$_2$-based thermal detectors.}) respectively. 

TeO$_2$ crystals have been established as a material with excellent performances as cryogenic calorimeters in CUORE \cite{Adams:2022, Adams:2024a} and its predecessors (MiBETA, Cuoricino, CUORE-0) \cite{ANDREOTTI2011822, Alduino_2016, Brofferio:2018}. CUORE has demonstrated the efficient scaling of bolometric experiments employing 750 kg of TeO$_2$ in 988 detectors. Despite the high radiopurity of the absorber material and of the detector's components used in the past and present TeO$_2$-based $0\nu\beta\beta$ experiments, the ROI is dominated by surface $\alpha$ radioactivity which contributes to a background level of about 10$^{-2}$~counts/(keV~kg~yr). This is caused by the almost identical response of a TeO$_2$ bolometer to $\alpha$ and $\beta$/$\gamma$ interactions of the same energy \cite{Bellini:2010}, which results to a lack of particle identification when using a single readout of thermal signals (more details about this approach can be found in \cite{Poda:2017}). It is worth noting that thanks to the high isotopic abundance of $^{130}$Te (about 34\% \cite{Meija:2016}), TeO$_2$ crystals with natural isotopic composition are and were used in CUORE and the predecessors. The only exception was a couple of $^{130}$Te-enriched ($^{130}$TeO$_2$) crystals additionally used in the MiBETA \cite{Pirro:2000} and Cuoricino \cite{Bryant:2010} experiments, showing poor performance and lower radiopurity in comparison with natural samples. Despite that, $^{130}$TeO$_2$ crystals are relevant for next-generation experiments like CUPID (CUORE Upgrade with Particle IDentification) \cite{Wang:2015raa}, simply because the enrichment would allow us to improve the half-life sensitivity of a TeO$_2$-based $0\nu\beta\beta$ experiment by a factor 3. In view of that, high performance radiopure $^{130}$TeO$_2$ bolometers (particularly, with the BINGO-/CUPID-like crystal size) have been recently developed \cite{Artusa:2017, CROSSenrichedTeO:2024}.

CUPID-Mo \cite{Armengaud:2020a, Augier:2022, Augier:2023} and CUPID-0 \cite{Azzolini:2018tum,Azzolini:2022} --- demonstrators of CUPID --- based on enriched Li$_2$$^{100}$MoO$_4$ and Zn$^{82}$Se (zinc selenide) scintillators respectively, have established and validated the performance of a new detector technique employing a heat-light readout to reject the surface $\alpha$ background, achieving the background level of $\bigO(10^{-3})$~counts/(keV~kg~yr) in ROI with few-ten kg scale array of thermal detectors \cite{Augier:2023, Azzolini:2019nmi}.
The heat-light readout is achieved thanks to an auxiliary light detector (LD), made of a Ge thin wafer, facing the main crystal and operated as a bolometer as well. For the same deposited energy, the light emitted after an $\alpha$ interaction in the main absorber is different compared to $\beta/\gamma$ interactions (typically, by a factor 5--10 lower for Li$_2$MoO$_4$ \cite{Pirro:2006,Tretyak:2010,Poda:2021}). This allows for a full $\alpha$ rejection in case of even poor scintillators like Li$_2$MoO$_4$, however, high-performance (i.e low-threshold) photodetectors are mandatory and this is one of the BINGO goals. The CUPID-Mo technique using Li$_2$MoO$_4$-based scintillating bolometers has been adopted for CUPID \cite{Armstrong:2019inu} (with 4.5-cm-side cubic crystals instead of cylindrical ones \cite{Armatol:2021a}), demonstrating the potential to achieve a projected background in the ROI at a level of approximately 10$^{-4}$ counts/(keV kg yr). A similar projection is expected for another next-generation experiment, AMoRE, which will use the same concept of Mo-containing scintillating bolometers, but with another type of a phonon sensor (metallic magnetic calorimeters) \cite{Alenkov:2015,Luqman:2017}. 

In the case of TeO$_2$, which exhibits only a tiny scintillation (not always detectable) \cite{Poda:2021}, we could rely only on the detection of the faint Cherenkov radiation \cite{Tabarelli:2010,Berge:2018}. The kinematic energy threshold of electrons and $\alpha$'s to produce Cherenkov light is $\sim$50~keV and $\sim$400~MeV respectively \cite{Tabarelli:2010}. Alpha particles from natural radioactive contaminants are by far too slow to emit Cherenkov radiation, therefore this could allow to distinguish them from $\beta/\gamma$ radiation. However, this sets very stringent requirements on the LD energy resolution ($\sim$20~eV RMS in optimal light collection conditions) and threshold as the collected Cherenkov signal from $\beta/\gamma$ interaction is of the order of only $\sim$100~eV per 2.6~MeV energy deposited in a CUORE-like ($5 \times 5 \times 5$~cm$^3$) large-volume crystal (see \cite{Poda:2021} and references therein). 
Once the background component of surface $\alpha$ radioactivity can be actively rejected with a TeO$_2$-based thermal detector array of CUORE-like scale, the background index according to the CUORE background model \cite{Adams:2024b} is predicted at $\bigO(10^{-3})$~counts/(keV~kg~yr), being dominated by the contribution of external sources of $\gamma$ quanta. The mitigation of this background issue can be achieved with a combination of highest radiopurity passive materials of the set-up together with a dedicated cryogenic veto (among the main BINGO goals), which could suppress by a factor 10 the background induced by the Compton scattering of 2615~keV $\gamma$ quanta of $^{208}$Tl (from the $^{232}$Th family of natural radioactivity).

Therefore, by exploiting the heat-light readout of thermal detectors in future $0\nu\beta\beta$ searches, one can expect to reject the dominant $\alpha$ background of thermal detectors and to reach the background in ROI at $\sim$10$^{-4}$~counts/(keV~kg~yr), in the best scenario. The further reduction of background in ROI to the level of $\sim$10$^{-5}$~counts/(keV~kg~yr) requires the rejection of surface-radioactivity-induced $\beta$ particles \cite{Artusa:2014a}. This can be achieved by innovative developments of an active background rejection of surface radioactivity (e.g. using a pulse-shape discrimination with metal-coated bolometers, which is under development within the CROSS project \cite{Bandac:2021}) and/or a new detector construction with minimized amount of passive materials (one of the main objectives of BINGO and some other current R\&Ds \cite{Alfonso:2022, Biassoni:2023, CROSSdetectorStructure:2024}).

\subsection{Innovative detector assembly}

As experienced by bolometric $0\nu\beta\beta$ experiments (e.g., the recent ones \cite{Alduino:2017, Azzolini:2019nmi, Augier:2023, Agrawal:2024, Adams:2024b}), the detector components play important contribution to the background budget in ROI. Therefore, the primary goal of the BINGO project is the development of a detector assembly with minimized amount of passive materials facing a crystal.  

\begin{figure}
\centering
\includegraphics[width=0.49\textwidth]{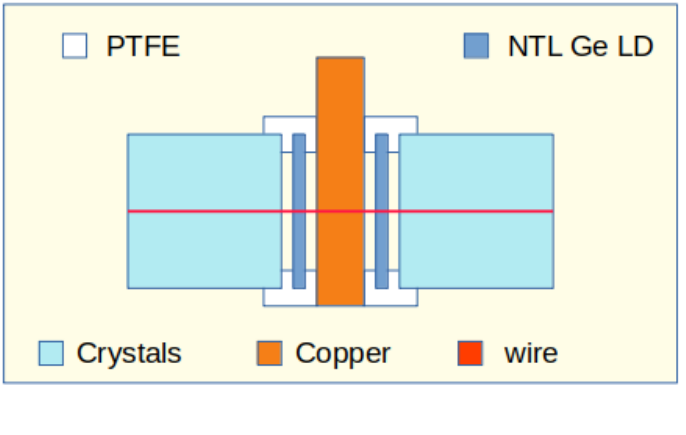}
\includegraphics[width=0.45\textwidth]{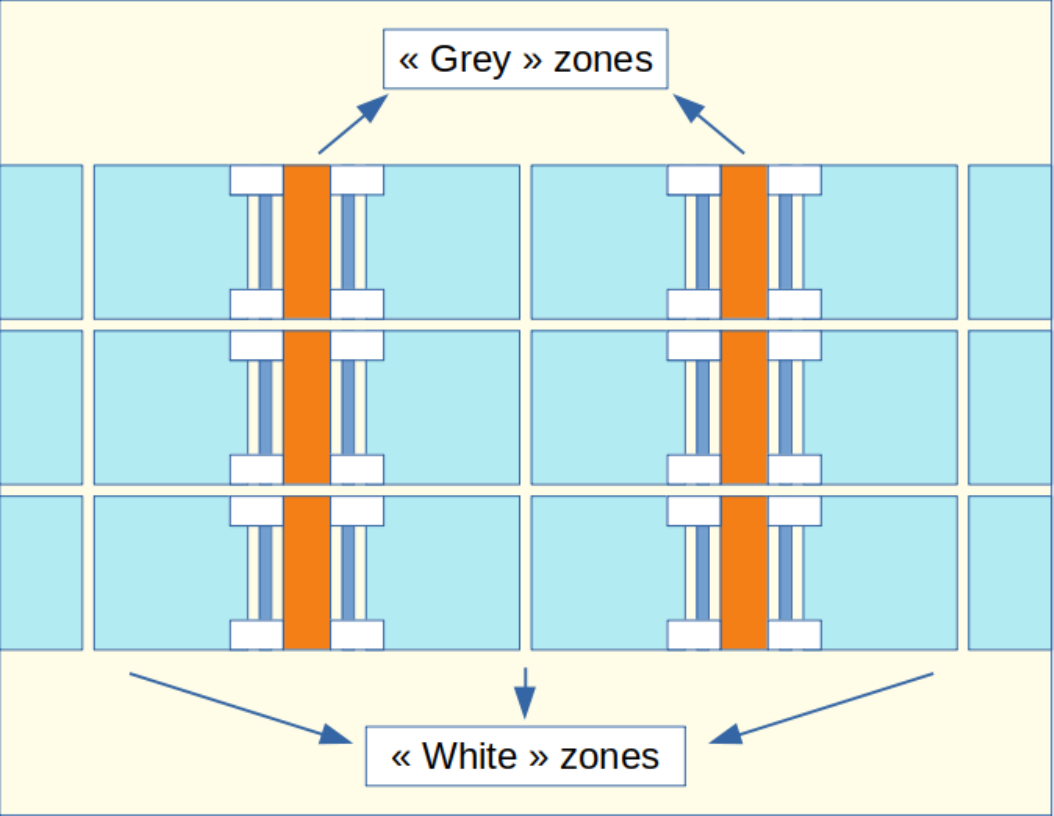}
\caption{Schematic view of the innovative BINGO assembly of a single module (\textit{left}) and a top part of the detector array (\textit{right}). A single module is comprised of four bolometers made of two cubic crystals and two square-shaped light detectors (labeled here as NTL Ge LD; see details in sec. \ref{sec:NTLLD}), in thermal contact to a Cu heat sink through supporting PTFE pieces and a wire. The array of BINGO modules has a large area of ``White'' zones with a small presence of passive materials around crystals (just PTFE elements and the wire), while the region with the highest concentration of the detector assembly components is shielded by LDs (``Grey'' zones), suppressing $\alpha$ and $\beta$ surface-induced radioactivity.}
\label{fig:BINGO_module_idea}
\end{figure}

BINGO proposes a new concept of mechanical assembly for heat-light bolometers, in which the main part of the detector holder is shielded by a vertically-placed LD. The main crystal, together with the LD Ge wafer and PTFE (polytetrafluoroethylene) supporting pieces, are hold by a wire under tension, as shown in Fig. \ref{fig:BINGO_module_idea} (\textit{left}). A single BINGO detector module is comprised of two crystals and two Ge LDs. Single modules can be arranged in towers forming a very compact array (providing an efficient multi-site-event rejection) with a significant part of ``White'' zones characterized by a small amount of passive materials, which are mainly concentrated in ``Grey'' zones, whose surface radioactivity can be efficiently screened by LDs, as illustrated in Fig.~\ref{fig:BINGO_module_idea} (\textit{right}). The development of the BINGO innovative assembly is the main objective of the present work and is described in details in the next sections (starting from Sec. \ref{sec:BINGO_assembly}). Preliminary results of GEANT4-based Monte Carlo simulations of a BINGO-like array in a CUORE-like set-up show a factor 10 improvement in the background in the ROI in comparison with a more classical approach of the detector assembly envisaged in CUPID \cite{Schmidt:2024}.

\subsection{Bolometric light detectors for BINGO}
\label{sec:NTLLD}

Considering the poor photon signal expected for the detector materials conceived in BINGO (as discussed above), the low photon collection efficiency in an open structure (i.e. without a reflective cavity) and the presence of a single LD per crystal, high-performance LDs are required to fulfil the BINGO background goals, based on particle identification capability.

One of the most viable technologies of high-performance bolometric LDs relies on the amplification of the light signal by the Neganov-Trofimov-Luke (NTL) effect \cite{NOVATI2018,Novati:2019}, which has a long-standing application in dark matter searches with massive thermal detectors based on semiconductors \cite{Pirro:2017}. 
This technology exploits the deposition of electrodes on the surface of a semiconductor absorber, which are then used to apply a voltage bias aiming to amplify the thermal signal. The drift of the initial electron-hole pairs, created by ionizing radiation, along the electric field lines induces an extra Joule heat proportional to the applied voltage (i.e. the NTL effect). An NTL LD which operates at 0 V electrode bias demonstrates similar performance as ``ordinary'' LD (i.e. without the deposited electrodes) \cite{Novati:2019,CrossCupidTower:2023a}. By applying a comparatively low voltage ($\sim$60--90~V), a factor 10 improvement in detector performance (e.g. signal-to-noise ratio, $S/N$) can be easily achieved, thus reducing the noise down to $\sim$10 eV RMS \cite{Novati:2019}. 

The technology of NTL LDs --- that is selected to be implemented in BINGO and recently chosen for CUPID --- is also essential to suppress the most critical source of background in $^{100}$Mo-enriched thermal detectors for $0\nu\beta\beta$ search \cite{Chernyak:2012,Chernyak:2014}: the random coincidences of events from the relatively fast $2\nu\beta\beta$ process ($T_{1/2} \approx 7\times10^{18}$~yr \cite{Augier:2023spectal}). Indeed, the $2\nu\beta\beta$ decay induces $\sim$10~mBq/kg activity in $^{100}$Mo-enriched bolometers \cite{Poda:2017}, which, in combination with their poor time resolution (ten(s) of ms), leads to the background in the ROI at the level of about 10$^{-2}$~counts/(keV~kg~yr) \cite{Chernyak:2012,Chernyak:2014}. The efficiency of the random coincidences rejection strongly depends on the detector's time response and $S/N$ \cite{Chernyak:2014}; for a CUPID-like massive bolometer (rise time $\sim$ 15 ms and $S/N$ $\sim$ 1000) this efficiency is around 99\% at the best \cite{Chernyak:2014,Armatol:2021b}. In CUPID, the contribution of this background source is required to be less than half of the total background budget of 10$^{-4}$~counts/(keV~kg~yr) \cite{Armstrong:2019inu}, which means the rejection of more than 99.7\% of random coincidences is needed. LDs, being small-mass bolometric devices, have faster rise-times ($\sim$1~ms or even less at high bias currents) compared to the Li$_2$MoO$_4$ heat channel, but a modest $S/N$ ($\sim$15) of the light channel makes it less efficient in pile-up rejection \cite{Chernyak:2014}. 
The NTL effect will improve substantially the $S/N$ of the Ge LDs currently used in bolometric experiments like CUPID-Mo and CUPID-0, and this will allow us to reject effectively the random coincidences using the light signal \cite{Chernyak:2017,Ahmine:2023}.

Finally, the NTL LD technology will be used in BINGO in the detection of scintillation signals from the crystals composing the cryogenic veto (described in the next section), to achieve the low-threshold required for efficient tagging of $\gamma$-induced background. 

An R\&D on development and optimization of NTL LDs for BINGO is ongoing. The main two tasks of this R\&D are the following: a) the development of an appropriate electrode design to improve the coverage of the Ge LD surface; b) the optimization of the fabrication method (i.e. the choice between shadow masks or photo-lithography to define the shape of the Al electrodes, which are deposited by electron-beam-assisted evaporation). Preliminary results on BINGO NTL LD developments are presented in \cite{Armatol:2023}.

\subsection{Cryogenic active veto}

\begin{figure}
\centering
\includegraphics[width=0.6\textwidth]{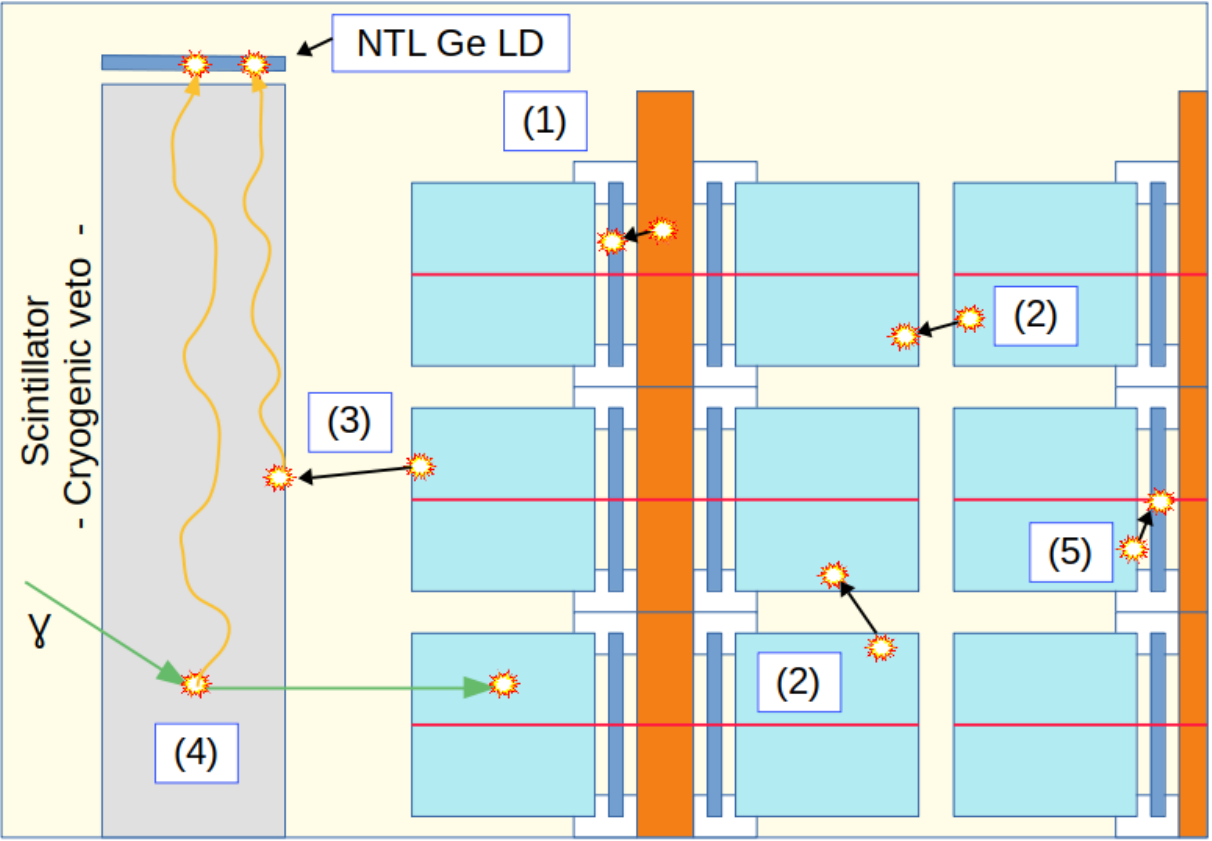}
\caption{Schematic view of a BINGO-like detector array with a cryogenic active veto based on a crystal scintillator cooled down to low temperatures and provided with a scintillation detection using Ge LDs with the NTL amplification. Different possible coincidences which can be exploited to reject external $\gamma$ and surface $\beta$ events thanks to BINGO improvements are labeled as follows: 
(1) Surface $\beta$s from the Cu holder can be rejected by coincidences with LDs, while surface $\alpha$ particles are completely screened by them; 
(2) Surface $\beta$s from crystals can be rejected by coincidences with another crystals; 
(3) Surface $\beta$s from crystals of the external layer of the array can be rejected by coincidences with the cryogenic veto;
(4) External $\gamma$s interacting with crystals can be rejected by coincidences with the veto; 
(5) Surface $\beta$s from the crystals can be rejected by coincidence with LDs.
}
\label{fig:BINGO_veto_idea}
\end{figure}

Once the suppression of the dominant background contribution in the ROI --- consisting of events generated by surface $\alpha$ and $\beta$ radioactivity from the detector components --- is achieved in a bolometric $0\nu\beta\beta$ experiment, $\gamma$ background can still mimic the effect searched for. Indeed, radioactive products of Th/U chains (e.g. $^{208}$Tl and $^{214}$Bi) emit $\gamma$ quanta with energies above $Q_{\beta\beta}$ and the Compton scattering of such $\gamma$s can be detected in the ROI, indistinguishable from the expected $0\nu\beta\beta$ event. A tag of the initial energy deposition of these high-energy $\gamma$ quanta, mostly originated from radioactive contamination in materials outside the experimental volume (i.e. external sources), is therefore mandatory to further reduce backgrounds in next-generation  $0\nu\beta\beta$ searches. For this purpose, BINGO proposes to use a cryogenic veto, as an active shield of the main detector (see in Fig. \ref{fig:BINGO_veto_idea}). 
The cryogenic veto can be made of an efficient crystal scintillator (large-volume elements) coupled to NTL LDs (to have a relatively fast time response and reasonably low detection threshold). The detection threshold of such veto has to be sufficient to tag a 50 keV energy deposition in the cryogenic scintillator and thus to reject efficiently a significant $\gamma$ background in $^{130}$Te ROI induced by the 2615 keV $\gamma$ of $^{208}$Tl, the end-point of the most intense natural $\gamma$ radioactivity. Moreover, the cryogenic veto can be useful to tag and reduce the residual crystal-surface-originated $\alpha$ and $\beta$ background from the external modules of the array, facing directly the veto crystals. Different background rejection capabilities via coincidences and screening available in the BINGO approach are illustrated in Fig.~\ref{fig:BINGO_veto_idea}.

Zinc tungstate (ZnWO$_4$) crystals were initially considered in the BINGO proposal for the construction of the cryogenic veto \cite{BINGO:ERC}. This choice was based on a high radiopurity \cite{Belli:2011a, Danevich:2018a, Belli:2019a} and a comparatively high light output of this scintillator at low temperatures \cite{Bavykina:2008, Kiefer:2016, Poda:2021}, as well as on the developed technology of a large-volume high-quality crystal growth \cite{Galashov:2009, Shlegel:2017, Belli:2022}. A relatively high scintillation has been also detected in the first BINGO-dedicated low-temperature measurement carried out with a 1 cm$^3$ ZnWO$_4$ crystal \cite{Dumoulin:2022}. In view of a faced technological limitation in the growth of rather long ZnWO$_4$ crystals foreseen in the BINGO cryogenic veto (around 30 cm of the cylindrical part), we pre-selected another inorganic scintillator, bismuth germanate (Bi$_4$Ge$_3$O$_{12}$), a well-known dense material for scintillation $\gamma$ detectors which can be grown with large enough diameter (about 13 cm) and length (around 40 cm) \cite{Borovlev:2001}. A comparative test of ZnWO$_4$ and Bi$_4$Ge$_3$O$_{12}$ samples with a size of $\oslash$3 cm $\times$ 6~cm each shows a twice higher scintillation signal emitted by the latter material \cite{Armatol:2022smallnylon}. A long cooling down of Bi$_4$Ge$_3$O$_{12}$ crystals, reported in previous measurements (e.g. \cite{Cardani:2012a}) and seen in our study \cite{Armatol:2022smallnylon}, leads to a longer time to be spent for the full detector thermalization down to the base temperature of the cryostat, but this is not problematic in a rare-decay experiment where long data takings are envisaged. Another known and more crucial issue of Bi$_4$Ge$_3$O$_{12}$ crystals is the presence of the radioactive $^{207}$Bi ($Q$-value = 2397 keV \cite{Wang:2017a}, $T_{1/2}$ = 31.2 yr \cite{Wang:2017b}) with a comparatively high activity (from hundreds mBq/kg to a few Bq/kg), which can be mitigated by pre-selection of raw materials used in the crystal production to get $^{207}$Bi activity in crystals $\sim$10 mBq/kg \cite{Danevich:2018a}.  

The first proof-of-concept of the BINGO cryogenic veto has been recently demonstrated with two Bi$_4$Ge$_3$O$_{12}$ trapezoidal bars (20 cm$^2$ $\times$ 12 cm), each coupled to two ``ordinary'' Ge LDs (i.e. no NTL amplification), and a TeO$_2$ crystal ($2\times 2 \times 2$ cm$^3$), which had a side facing the veto artificially contaminated with a U source to mimic the surface-induced radioactivity \cite{Khalife:2023}. An article on a final design of the BINGO cryogenic veto, to be used in a demonstrator experiment (detailed below), is in progress.

\subsection{MINI-BINGO demonstrator}
\label{sec:miniBINGO}

\begin{figure}
\centering
\includegraphics[width=0.7\textwidth]{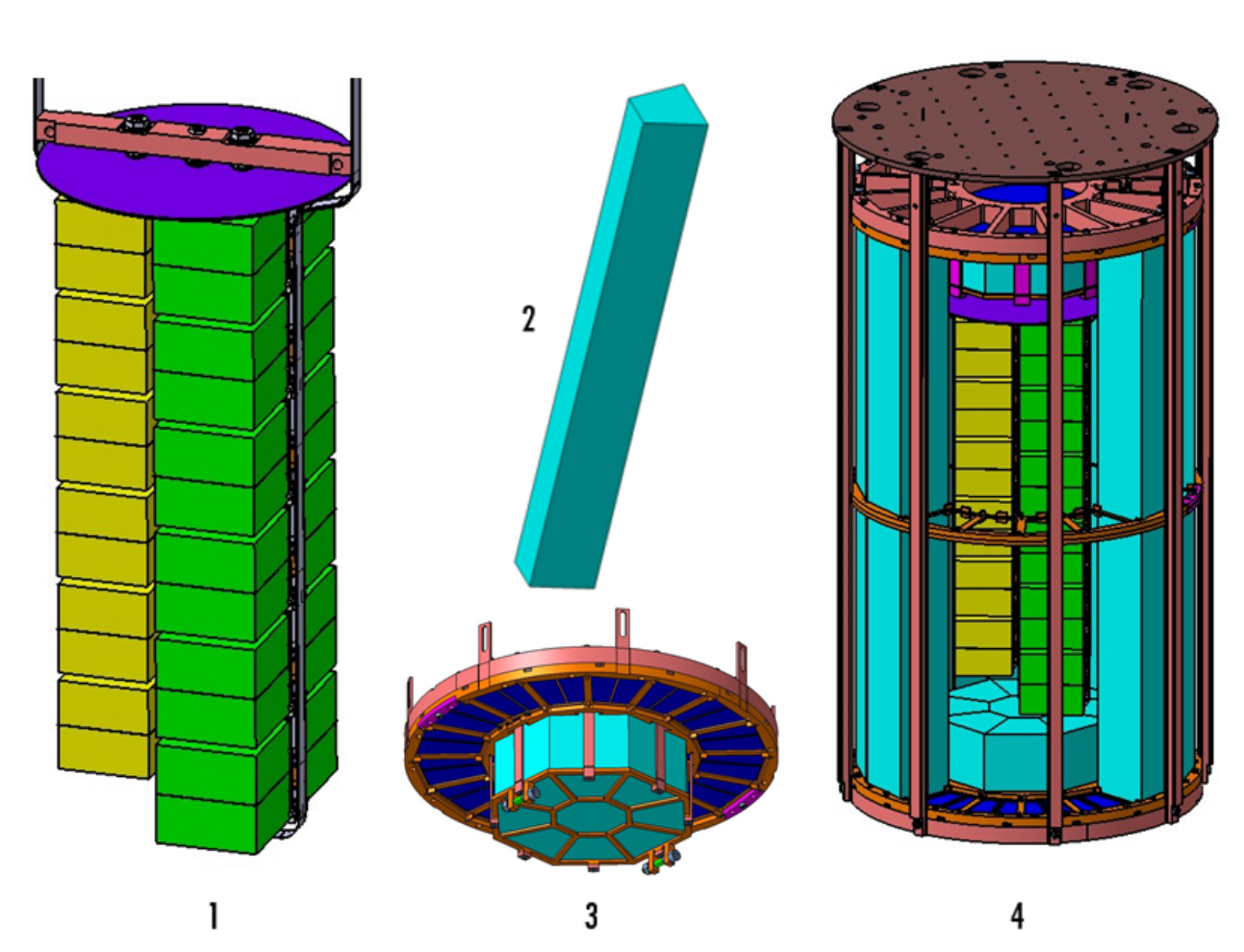}
\caption{Rendering of the MINI-BINGO detector components: 
(1) Towers of Li$_2$MoO$_4$ and TeO$_2$ cubic crystals with a side of 4.5 and 5.0 cm, respectively (each crystal is coupled to an NTL LD; not visible). A Si wafer ($\oslash$15 $\times$ 0.07 cm) is placed on the top of the tower as an additional detection element of the active shield; 
(2, 3) The active cryogenic veto, consisting of 16 Bi$_4$Ge$_3$O$_{12}$ bars on the lateral and two Bi$_4$Ge$_3$O$_{12}$-based modules on top and bottom. Each scintillator will be coupled to a NTL LD. 
(4) The full MINI-BINGO design showing the $0\nu\beta\beta$ detectors (towers) surrounded by the active veto, assembled in the Cu supporting structure (several Bi$_4$Ge$_3$O$_{12}$ bars are not shown for the internal set-up structure visibility). The size of the MINI-BINGO construction is around 30~cm in diameter and 50~cm in height; the total wight is around 170 kg (130 kg of crystals).}
\label{fig:scheme_mini_BINGO}
\end{figure}

All three above-presented BINGO innovations will be combined in a small-scale demonstrator --- MINI-BINGO --- which will consist of two towers (illustrated in Fig.~\ref{fig:scheme_mini_BINGO}): one tower with Li$_2$MoO$_4$-based bolometers (crystal size: $4.5\times 4.5 \times4.5$~cm$^3$) and another one with TeO$_2$ thermal detectors (crystal size: $5.1\times 5.1 \times5.1$~cm$^3$). Each tower will comprise 6 modules, each hosting 2 crystals. All crystals will be coupled to an NTL Ge LD with dimensions of $4.5\times 4.5 \times 0.03$~cm$^3$ for Li$_2$MoO$_4$ and $5.1\times5.1 \times 0.03$~cm$^3$ for TeO$_2$. The thickness of Ge wafers typically used in bolometric LDs is $\sim$0.005--1~mm \cite{Poda:2021} and it is less important for the detector's performance once the absorber mass remains at a gram-scale level (to have a low heat capacity). Thus, the thickness of BINGO LDs has been set based on the characteristics of commercial Ge wafers that meet our purity requirements. The towers will be surrounded by an active shield consisting of BGO scintillators coupled to NTL Ge LDs as well. The main components of the MINI-BINGO demonstrator are visualized in Fig. \ref{fig:scheme_mini_BINGO} and listed in Table \ref{tab:miniBINGO}.

\begin{table}
    \centering
    \caption{Principal components of the MINI-BINGO demonstrator. Each crystal of the MINI-BINGO towers will be coupled to a single NTL LD; each bar of the lateral veto will be viewed by two NTL LDs, while veto top / bottom arrays will be coupled to a single large ($\oslash$15~cm) NTL LD each.}   
    \vspace{8px}
    \begin{tabular}{|l|c|c|c|c|}
\hline
MINI-BINGO & Compound  & Number of & Crystal size  & Total mass \\ 
section & ~ & crystals & [shape] & (kg)  \\         
\hline
Molybdenum & Li$_2$MoO$_4$   & 12    & $4.5\times 4.5 \times 4.5$ cm$^3$    & 3.4 \\  
tower & ~   & ~    & [cubic]    & ~ \\  
\hline
Tellurium & TeO$_2$   & 12    & $5.1\times 5.1 \times 5.1$ cm$^3$    & 9.5 \\  
tower & ~   & ~    & [cubic]    & ~ \\  
\hline
Veto lateral & Bi$_4$Ge$_3$O$_{12}$ & 32 & 20 cm$^2$ $\times$ 23 cm & 105 \\  
~ & ~   & ~    & [trapezoidal]    & ~ \\  
\hline
Veto top  & Bi$_4$Ge$_3$O$_{12}$ & 8 & 18 cm$^2$ $\times$ 5.0 cm & 6.0 \\  
(Veto bottom) & ~   & ~    & [trapezoidal]    & ~ \\  
~ & ~   & 1    & 25 cm$^2$ $\times$ 5.0 cm   & ~ \\  
~ & ~   & ~    & [octagonal]    & ~ \\  
\hline
    \end{tabular}
    \label{tab:miniBINGO}
\end{table}

All 12 Li$_2$MoO$_4$ crystals have been already produced and 4 of them are investigated in the present work. The crystal production followed the protocol of purification and crystallization developed by LUMINEU \cite{Armengaud:2017,Grigorieva:2017} and applied for the batch production of $^{100}$Mo-enriched Li$_2$MoO$_4$ crystal scintillators for CUPID-Mo \cite{Armengaud:2020a} and CROSS \cite{Bandac:2020}, as well as a few $^{100}$Mo-depleted  Li$_2$MoO$_4$ samples \cite{CROSSdeplLMO:2023}. The TeO$_2$ tower will be filled with crystals previously used in the Cuoricino $0\nu\beta\beta$ experiment \cite{Alduino_2016}. 
A validation of capabilities of pre-selected producers of Bi$_4$Ge$_3$O$_{12}$ crystals to provide samples with the required size and radiopurity is ongoing.

The MINI-BINGO will be hosted in the Modane underground laboratory (LSM, France) in a low-background cryogenic set-up in 2025. A dedicated pulse-tube-based dilution refrigerator has been recently installed at the LSM and is under commissioning now \cite{Schmidt:2024}. 
A 1-yr long data taking is scheduled for the MINI-BINGO experiment to demonstrate the rejection efficiency of the backgrounds originated from the residual radioactive contamination in the detector and set-up. Such exposure would allow us to investigate the background of each tower with a rather high accuracy, compatible with the results of similar-scale $0\nu\beta\beta$ experiments with thermal detectors based on TeO$_2$ (Cuoricino, CUORE-0) and Li$_2$MoO$_4$ (CUPID-Mo, AMoRE-I) crystals.

\section{BINGO detector assembly}
\label{sec:BINGO_assembly}

\subsection{Design of the BINGO assembly}

\begin{figure}
\centering
\includegraphics[width=0.6\textwidth]{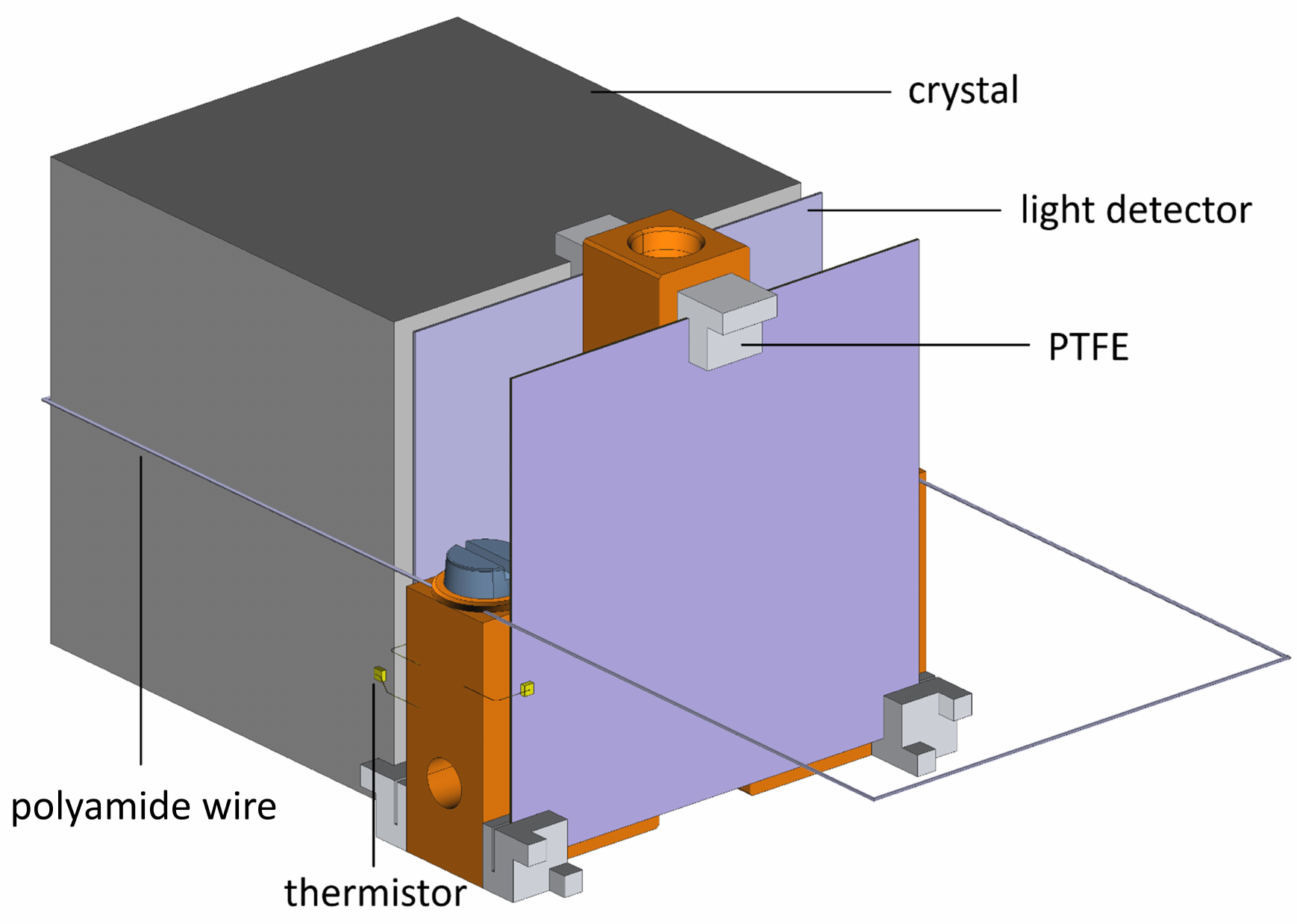}
\caption{A rendering showing the details of the BINGO assembly module. It consists of a single copper piece (in orange), three PTFE pieces per side to hold the light-detector Ge wafer and the crystal. A polyamide wire is used to fix the crystal against the PTFE pieces, which clamp the wafer. One of the two crystals is removed for a better visualisation of the assembly.}
\label{fig:scheme}
\end{figure}

As concluded from the previous section, an innovative detector structure forms the core of the BINGO project. The BINGO detector assembly, shown in detail in Fig.~\ref{fig:scheme}, is designed to fulfill three requirements: i) minimum passive materials facing the crystals; ii) open structure (no reflective foil around the crystals); iii) easiness of assembly. The remaining passive materials in the BINGO assembly are a copper support, a polyamide-based wire~\footnote{Here and the following we will use the expression ``polyamide wire'' for this supporting element of the BINGO module. The material used in the detectors here described is a commercial fishing line made of polyamide with a fluorocarbon coating (Caperlan~RESIST\textsuperscript{\textregistered}, Decathlon \cite{Decathlon_wire:2024}). Pure nylon can be used in the future if required, as this material can be obtained with extremely low concentration of radioactive impurities~\cite{Gando:2021}.} and PTFE pieces. 
The dominant part of the passive materials used in the detector assembly (Fig.~\ref{fig:scheme}) comprises a copper piece serving as the primary support for the crystals and the wafers, providing both mechanical and thermal connections to the cryostat. As in CUPID-0 and in CUPID-Mo, both the crystals and the Ge wafers are equipped with NTD (Neutron Transmutation Doped) Ge thermistors \cite{Haller:1994} for registering the thermal signal after particle or light interaction in the absorber. The copper piece works as a heat bath at temperatures around \textit{O}(10~mK) and serves as the anchoring point for the bonding pads, to which the thermistor readout wires are attached. 
The crystals are shielded from the copper support by the LDs, which are positioned vertically, effectively reducing backgrounds from residual surface radioactive impurities. Indeed, surface contamination on copper can produce electrons through a high $Q$-value $\beta$ decay (typically, $^{214}$Bi in the $^{238}$U chain) that can populate the $0\nu\beta\beta$ ROI in the energy spectrum. Those particles cannot be rejected via the heat-light combined detection. The remaining polyamide wire and PTFE elements cover about $\sim$1$\%$ of the crystal surfaces. They will be carefully selected and subjected to a strict cleaning and handling protocol to minimize their level of surface radioactivity. In an arrangement consisting of several BINGO towers, the five sides of the crystal not shielded by the LD will face directly another crystal rather than a passive element. 

The BINGO assembly design leads to a substantial reduction of passive surfaces --- by at least an order of magnitude --- compared to classical structures that are usually composed of copper frames and columns to host the crystals (see for example Ref.~\cite{Alduino_2016} for CUORE-0 and CUORE, and Ref.~\cite{Alfonso:2022} for CUPID). Furthermore, the resulting open structure allows us to efficiently identify crystal surface contamination through crystal-crystal coincidences and reject it through anti-coincidence cuts (as illustrated in Fig. \ref{fig:BINGO_veto_idea}).

\subsection{BINGO assembly: detector mounting and pre-testing}

\begin{figure}
\centering
\includegraphics[width=0.8\textwidth]{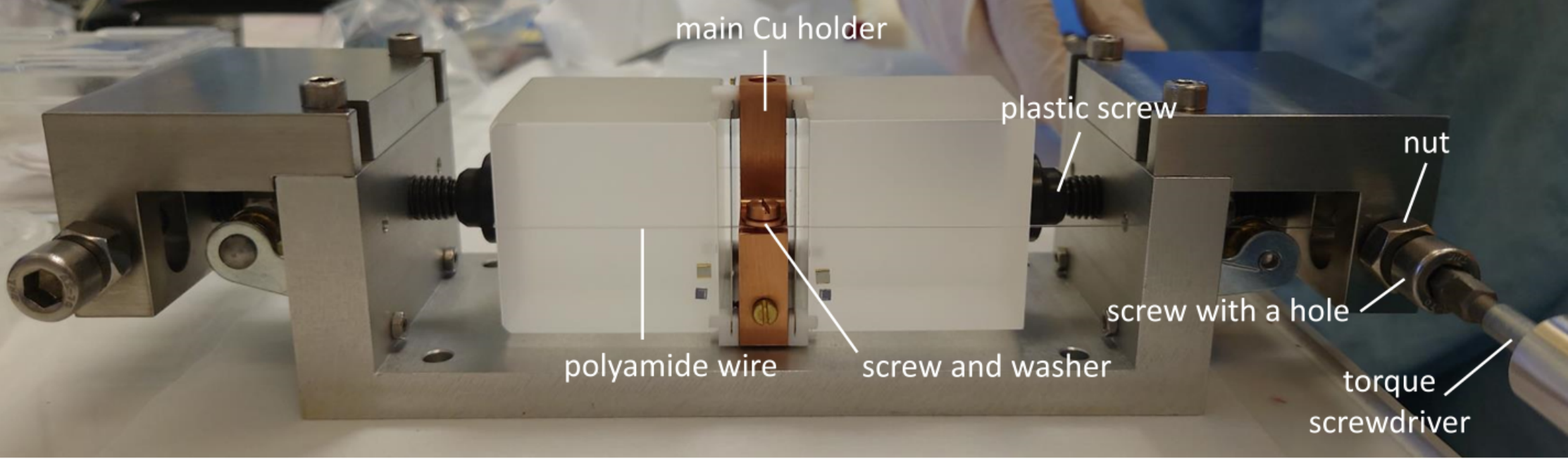}
\includegraphics[width=0.8\textwidth]{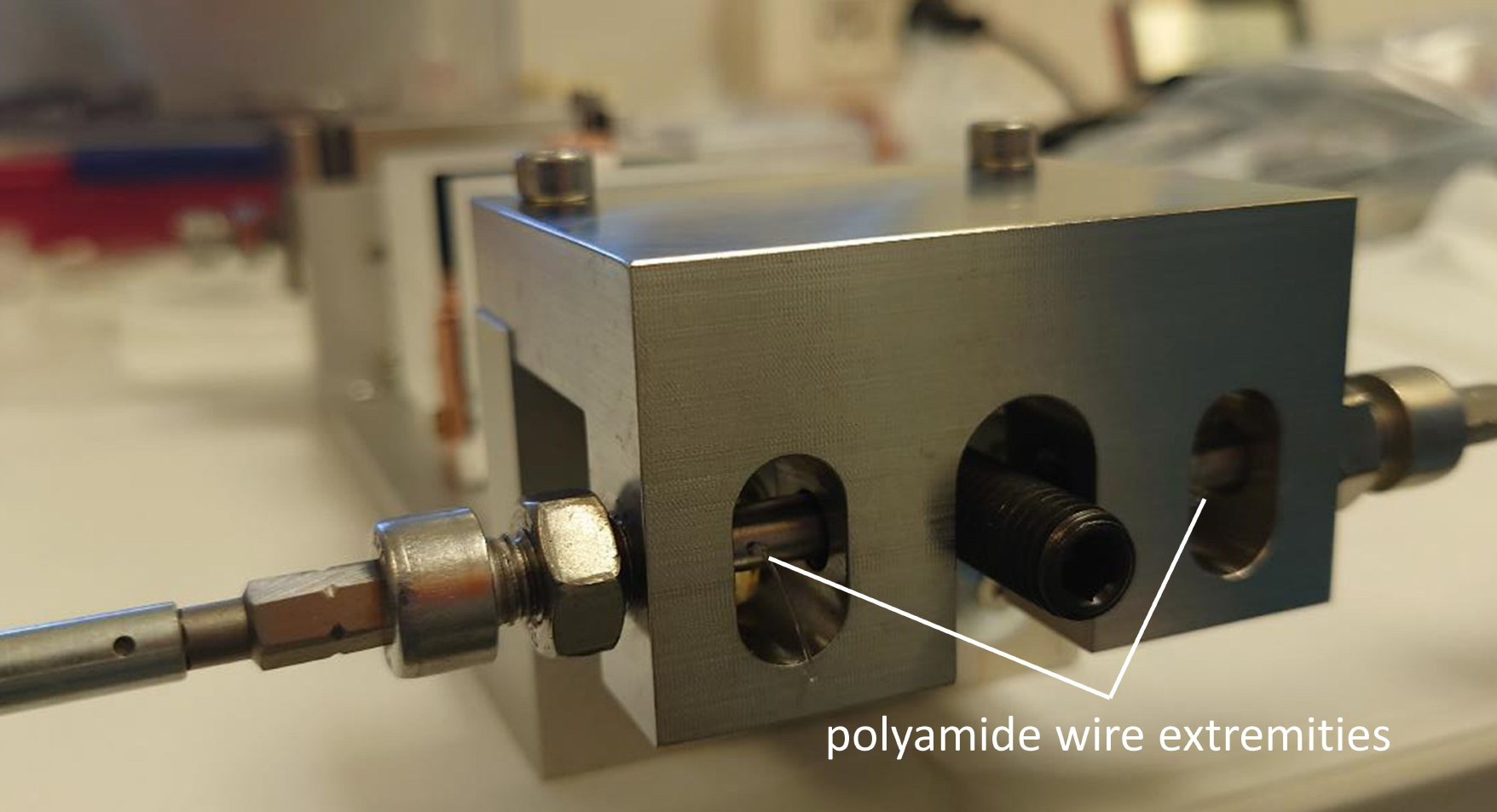}
\caption{\textit{(top)}: Single module placed in the mounting jig. The photo shows a polyamide wire around the left Li$_2$MoO$_4$ crystal to be fixed with the required tension using two torque screwdrivers for each wire end. \textit{(bottom)}: The two wire ends are pulled by passing them inside a hole made in the shaft of a screw that is subsequently rotated (violin string approach).}
\label{fig:assemliytool}
\end{figure}

We have developed and implemented a mounting jig (Fig.~\ref{fig:assemliytool}) to streamline the assembly process, including applying the necessary tension to the polyamide wire. This ensures efficiency, with a full two-crystal module assembly taking less than 5 minutes, and precision, as all elements are positioned with sub-millimeter accuracy. Initially, the aforementioned copper piece is secured to the mounting jig (Fig.~\ref{fig:assemliytool}, \textit{top}). Three PTFE pieces are affixed to the copper piece on each side, each featuring a 0.3~mm slot to securely hold the LD Ge wafer (Fig.~\ref{fig:scheme}). The crystal is positioned against these PTFE pieces and temporarily secured using a plastic screw, part of the mounting jig. A polyamide wire with a diameter of 0.45~mm, capable of withstanding a maximum tension of 16 kg, is wrapped around each crystal. A slot in the plastic screw allows us to position the polyamide wire around the crystal at mid-height and guide it under a copper washer, inserted in a corresponding copper screw. Each end of the polyamide wire passes through a small hole in the shaft of a winding screw, which can adjust the tension (Fig.~\ref{fig:assemliytool}, \textit{bottom}) by rotation, akin to tightening a violin or guitar string. Two torque-calibrated screwdrivers rotate the perforated screws, enabling us to apply a known tension (chosen as 4 kg on both sides) to pull the wire ends. Consequently, the crystal is pressed against the PTFE pieces, effectively fastening the Ge wafer. The 4 kg tension value was empirically determined as sufficient to prevent the Ge LDs from sliding between the PTFE pieces. The low thermal conductivity of PTFE and polyamide ensures thermal decoupling of both the Ge wafer and crystal from the heat bath. Most of the thermal conductance to copper occurs through the NTD Ge thermistors and the gold bonding wires for signal readout. The perforated screws are then secured from rotation using a nut. After fixing the second crystal in the same manner, the polyamide wires are anchored in place using the aforementioned copper screw and corresponding washer, compressing them against the copper holder. Subsequently, the wires are trimmed to their minimum necessary length, and the module is detached from the jig.

Consequently, each two-crystal BINGO module is assembled individually. The assembled modules are vertically stacked and held in place by a rod, similar to a skewer. This offers another advantage of the BINGO approach over CUORE and CUPID: the modules are fully independent, which can allow us to use devices constructed for crystal validation runs in a final experiment. Moreover, the easiness of the assembly procedure minimizes the likelihood of mistakes and failures and reduces handling, which in turn helps to control the radiopurity and mitigates the risk of exposure during the detector construction. 

Cooling down detectors to operation temperature will change the tension of the wires, primarily because the wire will contract much more than the crystal (typically, we expect an integral $\Delta L/L \sim -10^{-2}$ from room temperature to millikelvin range for polymers, to be compared with $\sim -10^{-3}$ for crystalline material)~\cite{Pobell:2007}. In addition, polyamide will change its physical properties. At very low temperatures, polyamide-based materials tend to experience an increase in brittleness and a decrease in ductility. This can result in reduced elongation at break. Tensile strength and modulus of elasticity may also be affected, typically decreasing as the temperature decreases. It is important to note that these trends are generalizations, and the specific mechanical properties of our assembly is impossible to predict in detail on the basis of the existing data. Since many physical properties change significantly already at the temperature of liquid nitrogen, we immersed modules made according to the previous design in this cryogenic fluid, subjecting them to repeated thermal cycles. We did not observe any detectable changes in the module structure or wire tension after these tests. This opened the way to test the bolometric performance of the BINGO detector modules in dilution refrigerators at millikelvin temperatures.

The first proof-of-concept of the assembly was performed on two $2.0\times2.0\times2.0$~cm$^3$ Li$_2$MoO$_4$ crystals operated as bolometers above-ground in a dry pulse-tube cryostat at IJCLab (Orsay, France) \cite{Armatol:2022smallnylon}. The characterization of small-volume (8~cm$^3$) Li$_2$MoO$_4$ crystals in the BINGO module showed a good bolometric performance that paved the way for testing the final BINGO size crystals (about 90~cm$^3$). The initial pre-test was carried out with a 2-crystal BINGO module operated above-ground, as reported in \cite{Armatol:2023largenylon}, while a long investigation in a more suitable background conditions (i.e in an underground low-background set-up) was performed with four large Li$_2$MoO$_4$ crystals, as described below.

\section{Construction and operation of a 4-crystal BINGO prototype}

\subsection{BINGO prototype with 4 large Li$_2$MoO$_4$ crystals}

For the construction and test of the first BINGO assembly with the large-size ($4.5\times 4.5 \times 4.5$~cm$^3$) Li$_2$MoO$_4$ crystals, we took randomly 4 samples from the batch of 12 crystals with natural molybdenum content produced for the MINI-BINGO demonstrator (see Sec. \ref{sec:miniBINGO}). As it concerns LDs, we used four  $4.5\times 4.5 \times 0. 3$~cm$^3$ Ge wafers (from the company Umicore) for the construction of ``ordinary'' LDs, with no Al electrode deposition (i.e. to be operated without the NTL mode). 
We would like to emphasize the importance of testing the initial performance of LDs (i.e., without the NTL signal amplification) in the BINGO assembly. This is crucial because the mechanical structure has no impact on the performance of the NTL amplification. We are confident that the existing technology of NTL LDs \cite{Novati:2019} is mature enough to provide the expected gain of $\sim$10 in $S/N$, while the currently ongoing R\&D on the electrode design \cite{Armatol:2023} would further enhance the performance of NTL LDs and these improvements can be easily implemented in MINI-BINGO.

\begin{figure}
\centering
\includegraphics[width = 0.5\textwidth]{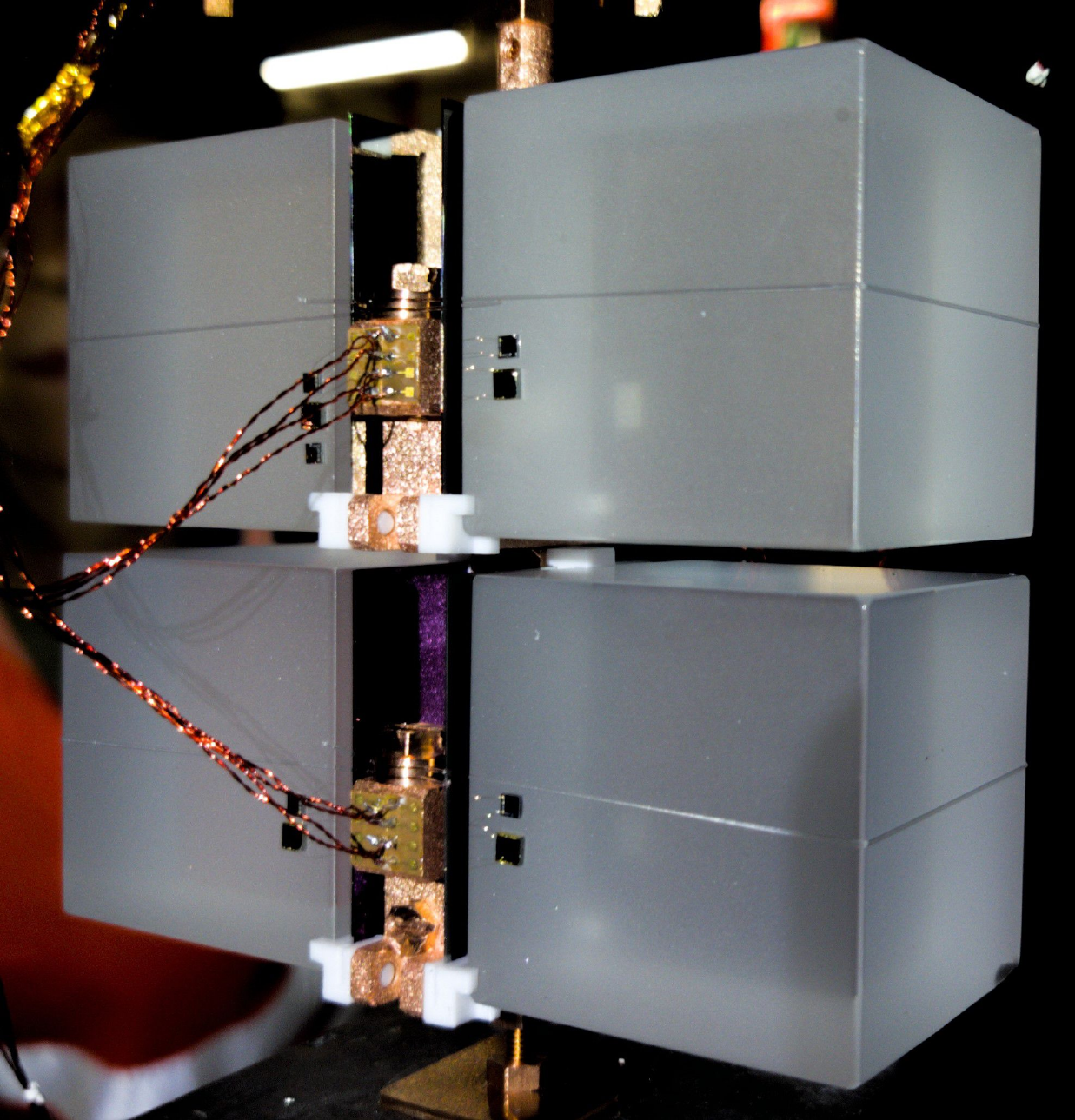}
\caption{Two BINGO modules, each hosting two Li$_2$MoO$_4$ crystals and two Ge light detectors.}
\label{fig:assemlymounted}
\end{figure}

In this assembly (Fig.~\ref{fig:assemlymounted}), one module had PTFE pieces and the other 3D printed PLA (polylactic acid) pieces to compare their impact on the bolometric performance. The advantage of 3D printing is that the shape and the size of the pieces can be changed quickly if required, but their radiopurity needs to be validated at least at the PTFE level~\cite{Alduino_2017}. The NTD Ge thermistors are glued manually to the crystal/wafer by means of 9 spots of two-component epoxy glue (Araldite\textsuperscript{\textregistered}~Rapid). A heavily doped Si chip \cite{Andreotti:2012} is also glued on each crystal, functioning as a heater to simulate a particle interaction. It is used for the monitoring and offline correction of thermal gain drifts of the bolometer response by delivering periodically a constant amount of energy \cite{Alessandrello:1998}. For the readout, NTDs and heaters are bonded using gold wires to pads on a Kapton\textsuperscript{\textregistered}~foil glued at the copper supporting piece.

\subsection{Detector operation}

The assembled two modules of the BINGO structure with four Li$_2$MoO$_4$ scintillating bolometers were operated underground in the Canfranc laboratory (LSC) in Spain. 
In the same run, six more ($^{100}$Mo-enriched/depleted) Li$_2$MoO$_4$ scintillating bolometers were assembled and tested (Run III in \cite{CROSSdetectorStructure:2024}), using holders designed for the CROSS experiment and consisting of copper and polymeric elements to fasten the crystals and the Ge wafers, as in CUORE, CUPID-0, and CUPID-Mo. The comparison with these CROSS detectors, which present the same crystal and wafer shape and size as BINGO, is useful to assess the performance of the innovative BINGO holder.
All the detectors were installed in the pulse-tube cryostat developed for the CROSS experiment \cite{Olivieri:2020, Ahmine:2023b}; a similar cryostat (the same producer) will be used in the MINI-BINGO demonstrator. The rock-overburden of 2450~m water equivalent, a 25-cm-thick external lead shield, an internal lead shield and a muon veto provide an excellent low radioactivity environment suitable for rare-event search campaigns with large bolometers. The facility furthermore employs an advanced magnetically damped vibrational decoupling system, which has proven particularly beneficial for the performance of bolometric LDs~\cite{Ahmine:2023b}. 

The detector readout was provided by a room-temperature front-end electronics consisting of low-noise DC-coupled voltage sensitive amplifiers, based on a Si JFET as input component \cite{Arnaboldi:2002}. The acquisition system consists of two 12-channel boards with integrated 24-bit ADC \cite{Carniti:2020, Carniti:2023}. The data were acquired with a sampling rate of 2~kS/s and low-pass Bessel with 300 Hz cut-off frequency. 

The facility is provided with room-temperature LEDs (880 nm) that allow us to deliver light to the experimental space by optic fibers that shine into the 10 mK cavity surrounding the detector. For the detector optimization, we inject pulses through the heaters and the LEDs to the Li$_2$MoO$_4$ bolometer and the Ge LD respectively, allowing us to find the optimum working point in terms of best signal-to-noise ratio.

During the experiment we regulated the detector plate temperature at 16~mK and performed a set of $\gamma$ calibrations with a $^{232}$Th source, and a series of background measurements. The $^{232}$Th source was inserted between the external lead shielding and the outer vacuum chamber of the cryostat. Depending on the position of the source with respect to our detectors, the intensity can vary because of the internal lead shield placed on the top of our experimental volume. Calibration data were acquired with high (close) and low (far) intensity $^{232}$Th source. The high intensity calibration allows us to increase the  $\gamma$ interaction rate in the Li$_2$MoO$_4$ crystals which leads to an emission of X-ray fluorescence of Mo at 17.5~keV that is used to calibrate the Ge LDs \cite{Armengaud:2020a}. However, large detectors can suffer from an increased pile-up rate degrading detector performance estimates. 

The detectors were cooled down for over 3 months in stable conditions. A possible variation of the crystal and wafer fastening strength could lead to an increase of the microphonic noise, given the high sensitivity of bolometers to vibrations. On the contrary, we observed a good stability of the bolometric performance and of the noise structure and level. This was confirmed by a careful room-temperature inspection of the assembly after the conclusion of the run, which showed no change in the detector arrangement.

A set of 251~h of $^{232}$Th calibration data, corresponding to the ones acquired at the optimized working point, was selected to study the bolometric performance of our detectors. Typical bias currents in the NTD thermistors were chosen of the order of 1~nA with operation resistances in the range 5--50~M$\Omega$. For the radiopurity measurement of the crystals (background measurements), around 609~h of data were selected and studied. The data processing is done offline using a MATLAB-based analysis tool \cite{Mancuso:2016} exploiting the Gatti-Manfredi optimum filter \cite{Gatti:1986}. The purpose of optimum filter is to maximise the signal-to-noise ratio in the signal bandwidth, providing the best estimation of the signal amplitude under a given template. The template is based on two inputs: an average mean pulse built using tens of high energy signals and an average noise power spectral density constructed from $\sim$10000 waveforms with no signals. The data are then triggered based on having an energy threshold above 5-RMS baseline-noise fluctuations. An n-tuple file is produced after data processing that contains a set of parameters characterizing our signals, including the uncalibrated amplitude of the filtered signal.

\section{Performance of the BINGO modules}

\subsection{Bolometric and spectrometric performance}

\begin{table}
    \centering
    \caption{Performance of Li$_2$MoO$_4$ bolometric detectors (rows 1--4) and Ge light detectors (rows 5--8) in the BINGO assembly obtained during a calibration with a $^{232}$Th source at 16~mK. The Li$_2$MoO$_4$ detector in row $i$ is coupled to the light detector in row $i+4$. The energy resolution of the Li$_2$MoO$_4$ bolometers at the 2615~keV $\gamma$ peak is quoted for a combined data with a high (88 h) and low (163 h) intensity of the source.}   
    \vspace{8px}
    \begin{tabular}{|l|c|c|c|c|c|}
        \hline
        Detector & Bias  & Resistance & Sensitivity & FWHM$_{bsl}$ & FWHM$_{2615}$ \\ 
        ~ & (nA) & (M$\Omega$) & (nV/keV) & (keV) & (keV) \\         
        \hline
        LMO-1   & 1.0   & 11    & 31    & 2.5 & 7.1 $\pm$ 0.4 \\    
        LMO-2   & 0.8   & 13    & 85    & 1.5 & 5.6 $\pm$ 0.2 \\    
        LMO-3   & 1.0   & 8.6   & 57    & 4.6 & 6.0 $\pm$ 0.4 \\    
        LMO-4   & 1.0   & 9.3   & 44    & 2.6 & 6.6 $\pm$ 0.4 \\    

        \hline
        LD-1    &1.0    & 5.8   & 985   & 0.24 &  -- \\     
        LD-2    &1.0    & 8.8   & 1660  & 0.16 &  -- \\     
        LD-3    &1.0    & 8.7   & 1790  & 0.21 &  -- \\     
        LD-4    &1.0    & 9.2   & 1320  & 0.26 &  -- \\     
        \hline
    \end{tabular}
    \label{tab:boloperf}
\end{table}

The bolometric performances are summarized in Table \ref{tab:boloperf} in terms of sensitivity and baseline resolution FWHM. The sensitivity provides the conversion between an energy deposition in the absorber and the corresponding sensor voltage variation usually given in nV/keV or $\mu$V/keV. The sensitivities of the BINGO detectors are typical and similar to those achieved by the devices in the CROSS holders \cite{CROSSdetectorStructure:2024}. The spread in sensitivity is expected to be related to non-uniformity in the coupling between the NTD and the crystal through the glue\footnote{It is worth noting that an automatic gluing system, developed at CEA for the CUPID experiment, will be exploited for the gluing of sensors onto the MINI-BINGO crystals.}. Concerning the baseline resolution of Li$_2$MoO$_4$, three crystals showed a good performance with 2.4~keV FWHM baseline width on the average, which is a good indicator of the intrinsic energy resolution. This value is in line with those observed in the best CROSS-holder detectors. The remaining crystal (LMO-3~ in Table \ref{tab:boloperf}) was affected by a higher noise that led to a baseline width about a factor two larger. As similar outliers in terms of baseline resolution have been observed with traditional assemblies (like CUPID-Mo) and we don't have any indication for a non-uniformity in the assembly of the module, we presently assume that this excess noise is likely explained by noise pick-up along the readout chain. Since the energy resolution of Li$_2$MoO$_4$ crystals at the $0\nu \beta \beta$ ROI (3034 MeV) is typically dominated by an energy dependent term \cite{Augier:2022}, this excess noise has a minimal impact on the $0\nu \beta \beta$ sensitivity, so we can quote an average resolution of all 4 detectors of 6.3 keV FWHM at 2615~keV ($^{208}$Tl calibration line). This resolution is slightly better than the harmonic mean resolution of 6.6 keV obtained for the full calibration data acquired in the CUPID-Mo $0\nu \beta \beta$ search \cite{Augier:2022} and is close to the CUPID goals \cite{Armstrong:2019inu}.

\begin{figure}
    \centering
		\includegraphics[scale=0.55]{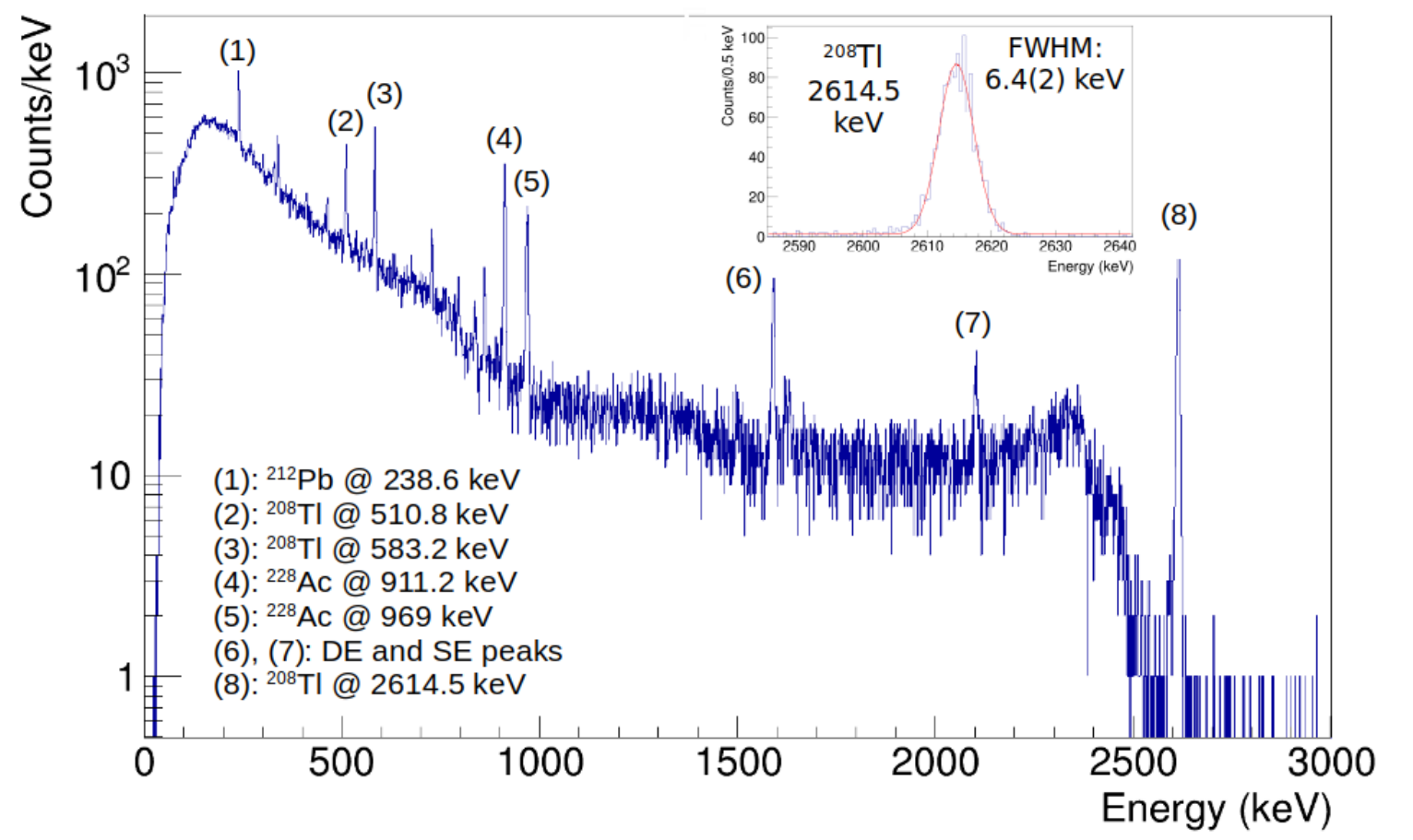}
    \caption{Energy spectrum obtained after 88 h of calibration with $^{232}$Th source (high intensity source mode) at 16~mK for all four Li$_2$MoO$_4$ bolometers.}
    \label{Fig:spectraLMO}
\end{figure}

\begin{figure}
    \centering
    \includegraphics[scale=0.3]{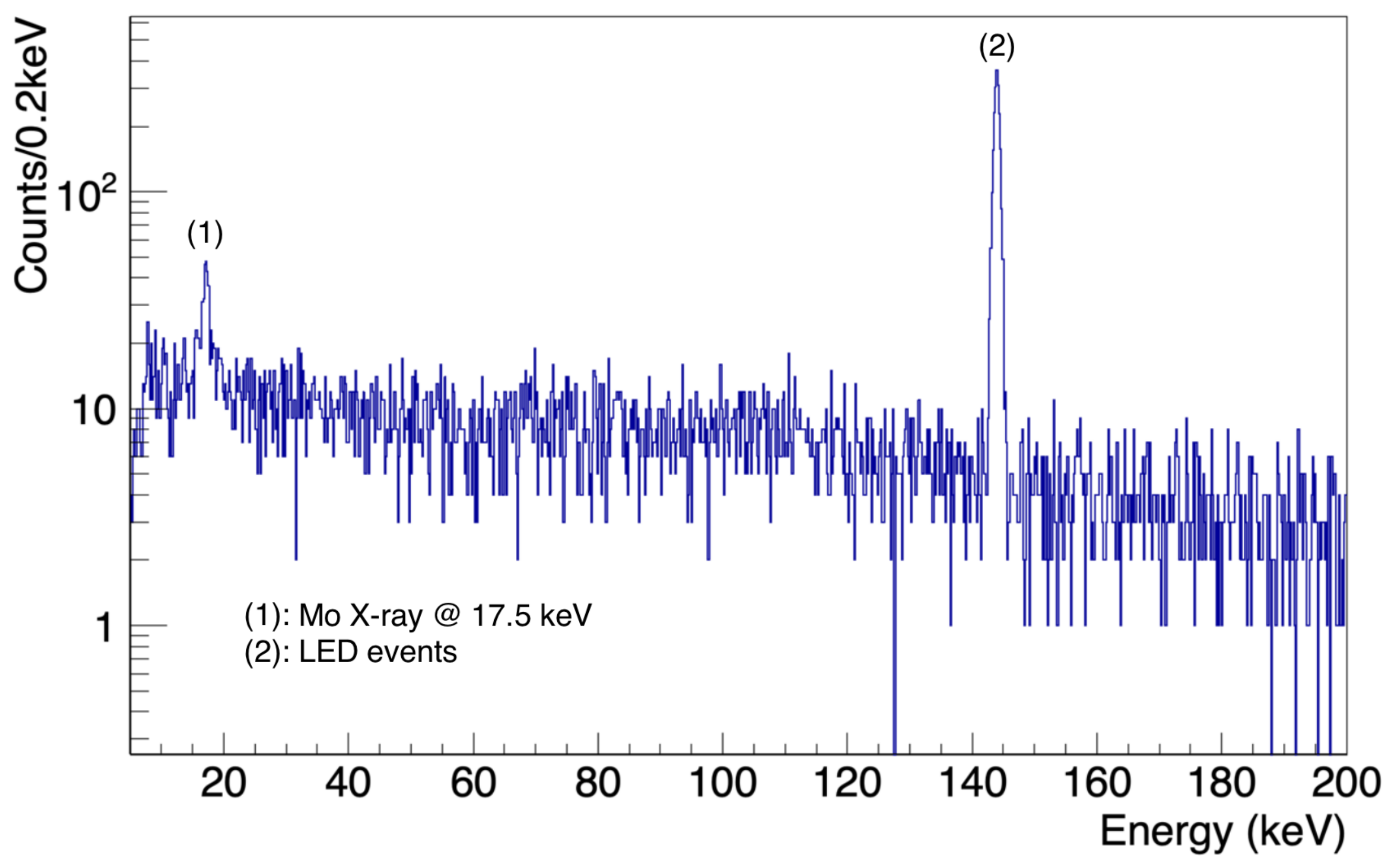}
    \caption{Energy spectrum obtained after 88 hours of calibration with $^{232}$Th source at 16 mK for a single Ge light detector (LD-2). A peak close to 20 keV is originated from X-rays of Mo induced in a coupled crystal (LMO-2) by $\gamma$ radioactivity, while the second peak at about 140 keV corresponds to LED signals injected for the monitoring of LDs operation.}
    \label{Figure:spectraLD}
\end{figure}

In terms of time structure of the pulses, we observe no difference with respect to the detectors housed in the CROSS holders within the typical spread of the pulse time constants. In particular, we have rise/decay times in the range 15--30~ms/50--120~ms for the heat channels (Li$_2$MoO$_4$) and 4--8~ms/13--23~ms for the light detectors. We remark that much faster rise-times can be obtained by raising the base temperature from 16~mK to $\sim 20$~mK and by increasing the bias current in the NTD Ge thermistors, with an acceptable loss in terms of energy resolution.

Thanks to the good energy resolution one can clearly see the thorium chain peaks used for the energy calibration of our detectors in Fig. \ref{Fig:spectraLMO}: the plot shows the combined high intensity source data which provides the best statistics in the low energy peaks, due to its unshielded position. In contrast, the low-intensity source data, with the source partially behind the lead shield, allowed us to assess pile-up effects in the 2615 keV peak. It provided compatible resolutions, ruling out any significant peak broadening and we quote combined detector performance values in Table~\ref{tab:boloperf}. 

As for the LDs, all of them work well in terms of sensitivity and baseline resolution, so one can expect a good particle discrimination. They have a baseline width in the range 70--110 eV  RMS (Table~\ref{tab:boloperf}).  Considering that a gain of a factor at least 10 in the signal-to-noise ratio is expected when the NTL mode is applied \cite{Novati:2019}, this performance can achieve not only full $\alpha$ rejection, but also suppression of pile-up events due to $2\nu\beta\beta$ compatible with a background index below 10$^{-4}$ counts/(keV~kg~yr) \cite{Ahmine:2023}, close to the level of what requested in next-generation bolometric $0\nu\beta\beta$ experiments. The energy spectrum of a single LD is shown in Fig. \ref{Figure:spectraLD}, with two apparent peaks of Mo X-rays at 17.5~keV and LED induced events.

We note also that there is no observable difference between PTFE (LMO-1 and LMO-4) and PLA (LMO-2 and LMO-3) assemblies as for their overall performances, which is an important result as PLA pieces are easier (and cheaper) to produce.

\begin{figure}
    \centering
    \includegraphics[scale=0.28]{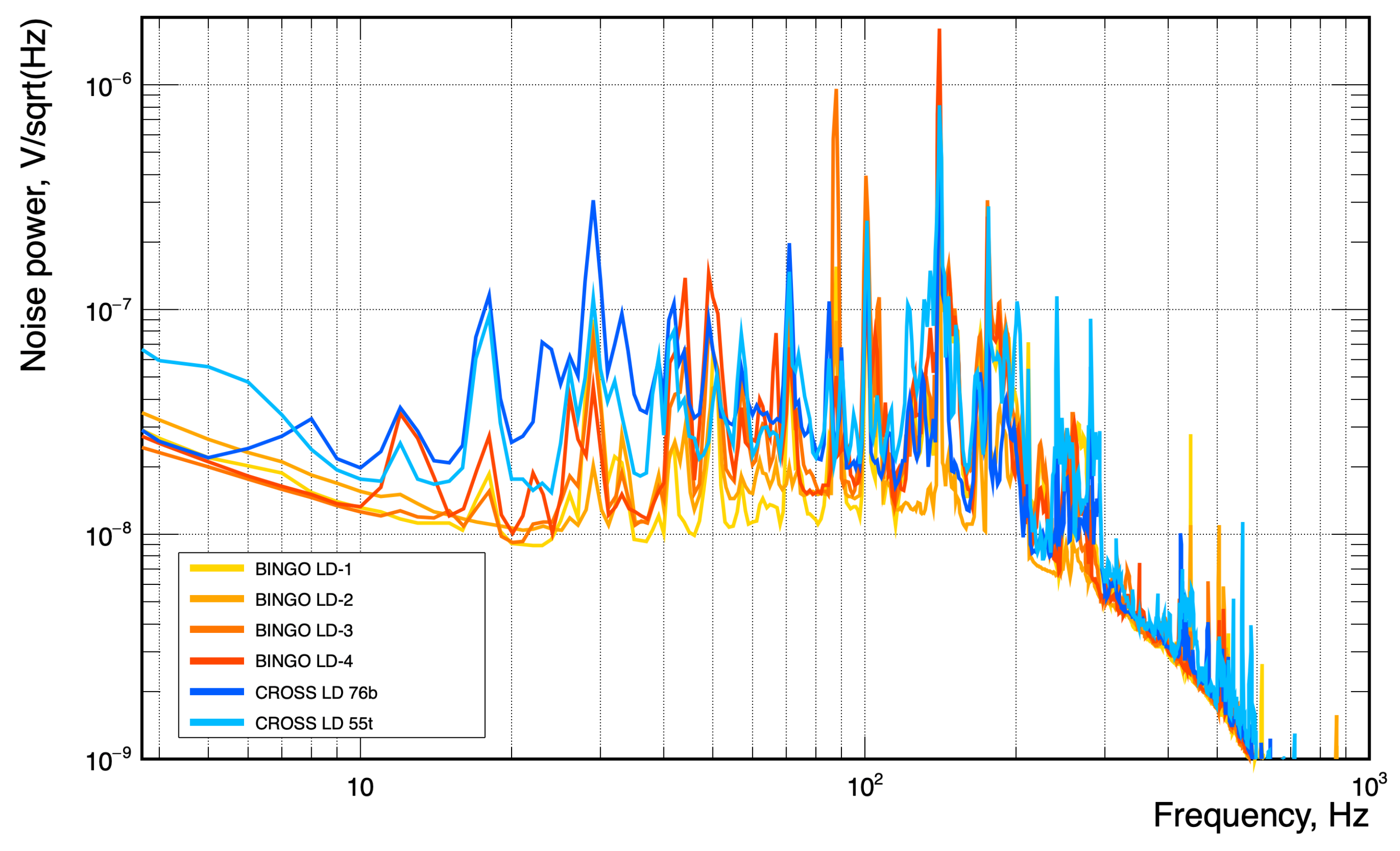}
    \caption{Comparison of the noise power spectra of the light detectors in the BINGO assembly (solid lines) with those in the CROSS assembly (dashed lines; Run III data from \cite{CROSSdetectorStructure:2024}) at 16 mK base temperature. The pulse tube of the cryostat is mainly responsible for the forest of the peaks present above 10 Hz \cite{Ahmine:2023}, with the channel-dependent impact \cite{CrossCupidTower:2023a}.} 
    \label{Fig:NPS}
\end{figure}

Since LDs are very sensitive to noise induced by vibrations, it is interesting to compare the noise power spectra of the LDs in the BINGO holder with those of the two best-performing detectors in the CROSS holder (Fig. \ref{Fig:NPS}). The figure shows that no additional specific noise structure is induced by the polyamide wire or any other BINGO assembly components. In more quantitative terms, we can compare the RMS voltage noise across the NTD Ge thermistors over a 5--50~Hz interval, which corresponds approximately to the useful bandwidth used by the optimum filter to provide the best signal-amplitude estimation. The integrated noise is 0.27~$\mu$V RMS (the spread is 0.17--0.42~$\mu$V RMS), averaged over the four BINGO LDs, to be compared with an average 0.56~$\mu$V RMS (0.4--0.7~$\mu$V RMS) for LDs in the CROSS structure. The small number of channels and the usual large spread in noise figures do not allow us to claim that the BINGO fastening method improves noise induced by vibrations, but we can surely conclude that it adds no relevant component.

\subsection{Particle identification with BINGO modules}

\begin{figure}
    \centering
		\includegraphics[scale=0.45]{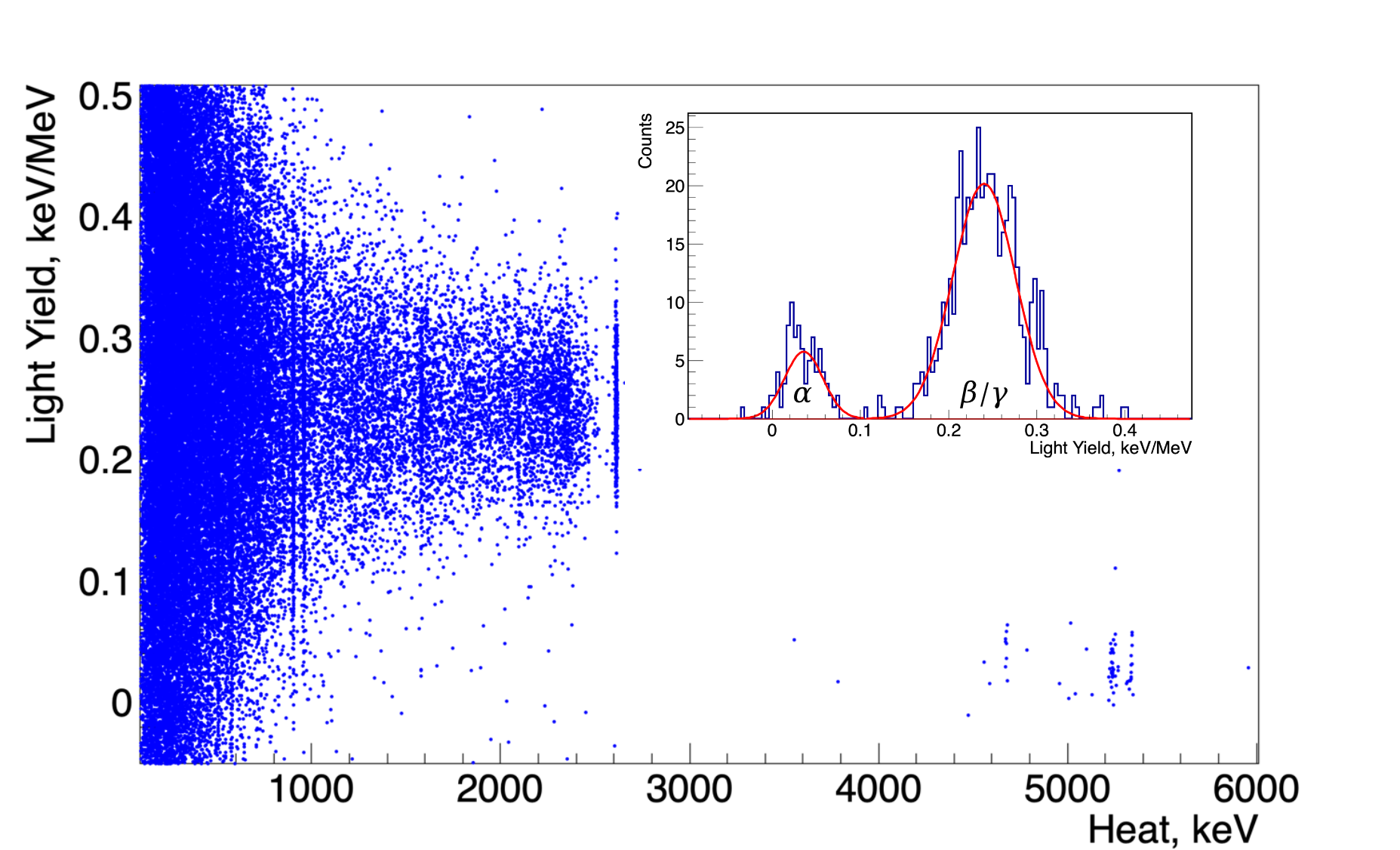}
    \caption{Light yield plot for LMO-2/LD-2 at 16 mK (Thorium calibration). The inset shows the Gaussian fit of the $\alpha$ and $\beta/\gamma$ distributions, where events were selected having an energy $> 2.5$~MeV. The estimated discrimination power is $\sim 5$.}
    \label{Fig:HL}
\end{figure}

For particle identification we use the light yield ($LY$) parameter, defined as the energy of the light signal detected by the Ge LD (measured in  keV after Ge LD energy calibration) for 1~MeV heat signal in Li$_2$MoO$_4$ (Fig. \ref{Fig:HL}). We point out that this parameter is not the absolute scintillation yield, as it depends also on the light collection efficiency. 
Table~\ref{tab:DP} summarizes the $LY$ of $\beta/\gamma$ events and the quenching (\textit{QF}) of the $\alpha$-particle scintillation light with respect to $\beta/\gamma$ for all detectors, where \textit{QF} = $LY_{\alpha}$ / $LY_{\beta/\gamma}$. We remark that LDs are not operated in the NTL mode.

The discrimination capability between $\alpha$ and $\beta/\gamma$ can be quantified in terms of the discrimination power (\textit{DP}):
\begin{equation}
    DP = \frac{|\mu_{\alpha}-\mu_{\beta/\gamma}|}{\sqrt{\sigma^2_{\alpha}+\sigma^2_{\beta/\gamma}}},
    \label{eq:DP}
\end{equation}
\noindent where $\mu$ and $\sigma$ are the mean and the standard deviation respectively extracted from the Gaussian fit applied to the $LY$ distributions of $\alpha$ and $\beta/\gamma$ events. The discrimination power was estimated from the $LY$ distribution corresponding to events higher than 2.5~MeV, close to our ROI. To achieve an $\alpha$ rejection higher than 99.9$\%$ (requested for a background level of $\sim 10^{-5}$ counts/(keV~kg~yr), a discrimination power higher than 3.2 is essential \cite{Berge:2018,Poda:2021}. In our case, two detectors over four showed a discrimination power 4--5, while the other two detectors around 3. This is a result of the worse baseline resolution in the corresponding LDs.

\begin{table}
    \centering
    \caption{Parameters of Li$_2$MoO$_4$ scintillation detection. We report on the relative light yield value for $\beta/\gamma$ events and the quenching factor of $\alpha$-particle induced scintillation light for events with energy $>$2.5~MeV. Uncertainties of light yield values are given (conservatively) as RMS values of the light yield distributions, while the uncertainties of mean values extracted from the fit are used to compute the uncertainties of the quenching factor.}   
    \vspace{8px}
    \begin{tabular}{|l|c|c|}
        \hline
        Detector & Light yield (keV/MeV) & Quenching factor ($\%$) \\ 
        \cline{2-3}
        ~ & $\beta/\gamma$ & $\alpha$ \\ 
        \hline
        LMO-1   & 0.22(3)   & 15(3)     \\
        LMO-2 & 0.25(3)   & 12(2)     \\
        LMO-3   & 0.24(4)   & 17(1)    \\
        LMO-4 & 0.22(4)   & 17(2)     \\
        \hline
    \end{tabular}
    \label{tab:DP}
\end{table}

The light yield achieved in the BINGO assembly is, as expected, lower compared to what can be achieved in a closed structure with a reflective foil around the crystals, as in CUPID-Mo ($LY_{\beta/\gamma} \sim$ 0.7~keV/MeV) \cite{Armengaud:2020a}, and similar to the open detector structure prototypes of CROSS and CUPID ($LY_{\beta/\gamma} \sim$ 0.2--0.3~keV/MeV) \cite{Armatol:2021a,Alfonso:2022,CrossCupidTower:2023a}. This drawback, however, will be exceedingly compensated by the use of the NTL effect which will result in an increase of $S/N$ by an order of magnitude for the Ge LD, leading to a much higher discrimination power.

\subsection{Radiopurity of the BINGO Li$_2$MoO$_4$ crystals}

To estimate the contamination levels in the tested Li$_2$MoO$_4$ crystals, we use the background measurements. This evaluation is important, as the four Li$_2$MoO$_4$ crystals belong to the production batch that will be used to assemble MINI-BINGO.

\begin{figure}
    \centering
     \includegraphics[scale=0.75]{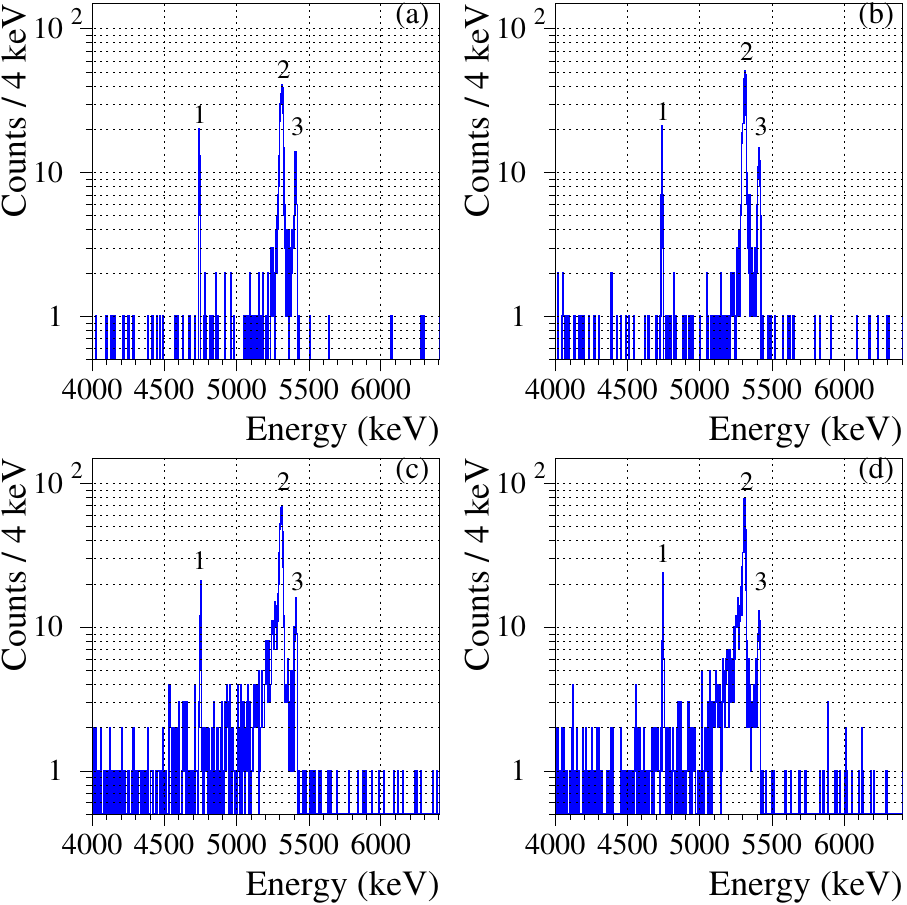} 
    \caption{Portions of the energy spectra in the $\alpha$ region of the following Li$_2$MoO$_4$ thermal detectors:  (a) LMO-1; (b) LMO-2; (c) LMO-3; (d) LMO-4. Data of all detectors were acquired over 609 h, except the LMO-1 bolometer (497 h), the NTD of which was grounded over a single dataset. The common prominent signatures, seen in the spectra, are labelled as follows: 1 --- products of $^{6}$Li(n,t)$\alpha$ reaction ($\alpha$ + triton) in the detector medium; 2 --- $\alpha$ particles emitted in $^{210}$Po decays occurred at surfaces of materials surrounding the detector; 3 --- products of $^{210}$Po decays ($\alpha$ + nuclear recoil) in the detector.} 
    \label{Fig:alpha}
\end{figure}

Using the $LY$ data (as illustrated in Fig.~\ref{Fig:HL}), $\alpha$ events were selected corresponding to a $LY$ below 0.15~keV/MeV and a heat energy above 3~MeV. The energy spectra of the $\alpha$ events detected by the Li$_2$MoO$_4$ bolometers are shown in Fig.~\ref{Fig:alpha}. The spectra were calibrated using the $\alpha$ peaks to account for their thermal quenching \cite{Armengaud:2017}. 

Three peaks are visible in Fig.~\ref{Fig:alpha}. The 4.7~MeV line corresponds to the neutron capture on $^6$Li resulting in an emission of an $\alpha$ and a triton: $^{6}$Li(n,t)$\alpha$. The two higher energy peaks correspond to the $Q$-value and $\alpha$-peak from $^{210}$Po decay to $^{206}$Pb. The 5.4~MeV line ($Q$-value) is for bulk $^{210}$Po contamination where all the emitted $\alpha$ and recoil nucleus energies are absorbed in the crystal. The lower energy $\alpha$'s at around 5.3~MeV correspond to surface $^{210}$Po contamination of the detector components, where the nuclear recoil energy is not absorbed in the Li$_2$MoO$_4$ thermal detectors.  

The average $^{210}$Po internal activity was estimated to be 124(8)~$\mu$Bq/kg, consistent with what was found previously for Li$_2$MoO$_4$ crystals \cite{Armengaud:2017, Augier:2023}. (The event selection efficiency, about 96\% for each detector, was estimated with heater-induced events.) Taking into account that we do not find a clear evidence of $^{226}$Ra and $^{228}$Th, we took all events in 50 keV energy interval centered around the corresponding $Q$-value as a signal, while the background was estimated in the energy range free of natural $\alpha$ radionuclides, as described in \cite{Armengaud:2015}. Following the Feldman-Cousins procedure \cite{Feldman:1998}, we calculated the number of events of the $\alpha$ peaks searched for excluded at a given confidence level (90\% C.L.). Thus, we set upper limits on $^{226}$Ra and $^{228}$Th activities at the level of 10~$\mu$Bq/kg and we expect that the contamination will likely be more than an order of magnitude below these values as found in CUPID-Mo \cite{Augier:2023}. The results of the study of $\alpha$~spectra are summarized in Table~\ref{tab:alpha_rate}.

\begin{table}
    \centering
    \caption{Results of the analysis of alpha events detected by Li$_2$MoO$_4$ bolometers. We report the rate of $\alpha$+t events (from neutron capture on $^6$Li) as well as estimates of activities of radionuclides from U/Th chain; uncertainties are given at 68\% C.L., while limits are set at 90\% C.L.}    
    \vspace{8px}
    \begin{tabular}{|l|c|c|c|c|}
        \hline
        Detector & Rate (counts/d)   & \multicolumn{3}{c|}{Activity ($\mu$Bq/kg)}\\ 
        \cline{2-5}
        ~       & $\alpha$+t & $^{210}$Po & $^{226}$Ra & $^{228}$Th \\
        \hline
        LMO-1   & 2.5 $\pm$ 0.4   & 137 $\pm$ 17 & $\leq$ 5  & $\leq$ 9   \\
        LMO-2   & 2.2 $\pm$ 0.3   & 129 $\pm$ 15 & $\leq$ 8  & $\leq$ 8   \\
        LMO-3   & 2.1 $\pm$ 0.3   & 117 $\pm$ 14 & $\leq$ 10 & $\leq$ 10   \\
        LMO-4   & 2.4 $\pm$ 0.3   & 113 $\pm$ 14 & $\leq$ 13 & $\leq$ 10   \\
        \hline
    \end{tabular}
    \label{tab:alpha_rate}
\end{table}

\noindent

\section{Conclusions}

We present an overview of the BINGO project and its R\&D program aiming at the development of innovative technologies for background reduction in next-generation experiments with thermal detectors to search for $0\nu\beta\beta$ decay in $^{100}$Mo and $^{130}$Te. BINGO focuses on the three main developments: a) a new detector structures with a minimal amount of passive materials facing crystals directly; b) high-performance bolometric Ge light detectors with Neganov-Trofimov-Luke signal amplification; c) a cryogenic veto with the scintillation readout based on bolometric light detectors. The efficiency of the background mitigation techniques proposed by BINGO will be demonstrated in a small-scale experiment (MINI-BINGO), to be started in a pulse-tube-based set-up at Modane underground laboratory (France) in 2025. MINI-BINGO will host 12 lithium molybdate (Li$_2$MoO$_4$, 4.5-cm side cubes) and 12 tellurium dioxide (5.1-cm side cubes) crystals, accompanied with thin Ge light detectors and surrounded by a 5-cm-thick cryogenic veto based on large bismuth germanate crystals (32 23-cm-long trapezoidal bars on the lateral side and 9 smaller crystals with octagonal arrangement on the top / bottom sides).

We describe in detail the development of the core of the BINGO project --- the innovative detector assembly with a wire-based support --- and its validation in a low-temperature measurement. In particular, we demonstrate the effective bolometric performance of the BINGO assembly tested with two modules containing 4 Li$_2$MoO$_4$ crystals ($4.5 \times 4.5 \times 4.5$~cm$^3$ each), taken randomly from the batch of samples produced for MINI-BINGO, and 4 Ge light detectors ($4.5 \times 4.5 \times 0.03$~cm$^3$). The BINGO detector structure offers stable mechanical coupling, and the detectors exhibit noise levels comparable to those in more conventional mounting structures. This finding holds significant implications for future $0\nu\beta\beta$ bolometric experiments, as it introduces an innovative assembly structure that drastically reduces the surface of passive materials facing the crystals by at least an order of magnitude. Consequently, this will lead to a substantial reduction in the radioactive background within the region of interest originating from surrounding materials. 
Thanks to a comparatively long operation (about 25 d) of the 4-crystal BINGO detector prototype in background measurements, we demonstrate a high radiopurity of the Li$_2$MoO$_4$ crystals (activities of $^{228}$Th and $^{226}$Ra are below 0.01 mBq/kg) to be used in MINI-BINGO. 
Prior to finalizing the MINI-BINGO towers, we will conduct additional prototype tests using $5.1\times5.1\times5.1$~cm$^3$ TeO$_2$ crystals, which are heavier, to validate the performance of the tellurium section of BINGO. Simultaneously, ongoing simulations aim to quantify the potential background reduction achievable through this novel assembly method  (preliminary results can be found in \cite{Schmidt:2024}).

\section*{Acknowledgements}

This work is supported by the European Research Council (ERC) under the European Union’s Horizon 2020 research and innovation program (Project BINGO, Grant agreement ID: 865844). 
We thank the CROSS collaboration for providing the underground cryogenic facility and the detector readout system. CROSS is also supported by the ERC in the framework of Horizon 2020 (Grant No. ERC-2016-ADG, ID 742345). The INR NASU group was supported in part by the National Research Foundation of Ukraine (Grant No. 2023.03/0213).

\end{document}